\documentclass[useAMS,usenatbib,babel,times]{mn2e}
\usepackage{multirow,balance,verbatim}
\usepackage[english,english]{babel}
\usepackage{amsmath,subfig}
\usepackage{amssymb,amsfonts,textcomp}
\usepackage{array}
\usepackage{supertabular}
\usepackage{hhline}
\usepackage{soul}
\usepackage{hyperref}
\hypersetup{dvips, colorlinks=true, linkcolor=black, citecolor=black, filecolor=black, urlcolor=black}
\usepackage[dvips]{graphicx}
\def\gtrsim{\lower.5ex\hbox{$\; \buildrel > \over \sim \;$}}
\usepackage{graphicx}
\usepackage[a4paper, total={7in, 9in}]{geometry}
\usepackage[usenames]{color}

\def \apj {{\it ApJ}}
\def \apjl {{\it ApJ Let.}}
\def \apjs {{\it ApJ Sup.}}

\def \aap {{\it A\&A}}

\def \mnras {{\it MNRAS}}
\def \nat {{\it Nature}}
\def \procspie {{\it Proc.~SPIE}}

\def\jcap{{\it JCAP}}
\def\physrep{{\it Phys.~Rep.}}

\begin{document}

\author[Chisari et al.]{
  \parbox{\textwidth}{N. E. Chisari$^1$\thanks{elisa.chisari@physics.ox.ac.uk},
    M. L. A. Richardson$^{1,2}$, J. Devriendt$^1$, Y. Dubois$^3$, A. Schneider$^4$, A. M. C. Le Brun$^{5,6}$, R. S. Beckmann$^{1,3}$, S. Peirani$^{3,7}$, A. Slyz$^1$, C. Pichon$^{3,8}$}
\vspace*{6pt}\\
\noindent
$^{1}$Department of Physics, University of Oxford, Keble Road, Oxford, OX1 3RH,UK.\\
$^{2}$Department of Astrophysics, American Museum of Natural History, 79th Street at Central Park West, New York, NY 10024, USA. \\
$^{3}$Institut d'Astrophysique de Paris, CNRS \& UPMC, UMR 7095, 98 bis Boulevard Arago, 75014, Paris, France.\\
$^{4}$Institute for Particle Physics and Astrophysics, ETH Zurich, Wolfgang-Pauli-Strasse 27, CH-8093 Zurich, Switzerland.\\
$^{5}$IRFU, CEA, Universit\'e Paris-Saclay, F-91191 Gif-sur-Yvette, France.\\
$^{6}$Universit\'e Paris Diderot, AIM, Sorbonne Paris Cit\'e, CEA, CNRS, F-91191 Gif-sur-Yvette, France.\\
$^{7}$Universit\'e C\^ote d'Azur, Observatoire de la C\^ote d'Azur, CNRS, Laboratoire Lagrange, France.\\
$^{8}$Korea Institute of Advanced studies (KIAS), 85 Hoegiro, Dongdaemun-gu, Seoul, 02455, Republic of Korea.\\
}

\date{Accepted 2018 July 30. Received 2018 July 26; in original form 2018 January 25.}

\title[Baryons \& the matter power spectrum]{                                                                
The impact of baryons on the matter power spectrum from the Horizon-AGN cosmological hydrodynamical simulation}

\maketitle

\begin{abstract}

Accurate cosmology from upcoming weak lensing surveys relies on knowledge of the total matter power spectrum at percent level at scales $k < 10$ $h$/Mpc, for which modelling the impact of baryonic physics is crucial. We compare measurements of the total matter power spectrum from the Horizon cosmological hydrodynamical simulations: a dark matter-only run, one with full baryonic physics, and another lacking Active Galactic Nuclei (AGN) feedback. Baryons cause a suppression of power at $k\simeq 10$ $h/$Mpc of $<15\%$ at $z=0$, and an enhancement of a factor of a few at smaller scales due to the more efficient cooling and star formation. The results are sensitive to the presence of the highest mass haloes in the simulation and the distribution of dark matter is also impacted up to a few percent. The redshift evolution of the effect is non-monotonic throughout $z=0-5$ due to an interplay between AGN feedback and gas pressure, and the growth of structure. We investigate the effectiveness of an analytic ``baryonic correction model'' in describing our results. We require a different redshift evolution and propose an alternative fitting function with $4$ free parameters that reproduces our results within $5\%$. Compared to other simulations, we find the impact of baryonic processes on the total matter power spectrum to be smaller at $z=0$. Correspondingly, our results suggest that AGN feedback is not strong enough in the simulation. Total matter power spectra from the Horizon simulations are made publicly available at \url{https://www.horizon-simulation.org/catalogues.html}.
\end{abstract}

\begin{keywords}
cosmology: theory ---
gravitational lensing: weak --
large-scale structure of Universe ---
methods: numerical
\end{keywords}

\section{Introduction}
\label{intro} 

The next generation of optical galaxy surveys will rely on different proxies for the distribution of matter in the Universe to constrain its components and their evolution. One such observable is the weak gravitational lensing of galaxies, percent-level distortions of their ellipticities which are caused by bending of the path of distant photons due to gravity, first detected by \citet{Tyson90}. These distortions can be used to map the distribution of matter in the Universe through cosmic time to set constraints on the evolution of dark energy, theories of gravity and the nature of dark matter, among other applications. For reviews of weak lensing theory, methods and applications, see \citet{BS01,Weinberg13,Kilbinger15}.

Previous and ongoing weak lensing surveys, such as the Sloan Digital Sky Survey \citep{Huff14}, the Canada-France-Hawaii Telescope Lensing Survey \citep{Heymans12}, the Kilo Degree Survey \citep{deJong13}, the Deep Lens Survey \citep{Wittman02} and the Dark Energy Survey \citep{Troxel17}, among others, have demonstrated the feasibility and potential of this method for precision cosmology. Future experiments with better constraining power are planned or under construction, such as the Large Synoptic Survey Telescope \citep[][LSST]{LSST}, {\it Euclid} \citep{Laureijs11} and {\it WFIRST} \citep{green11}. To successfully extract information from weak gravitational lensing measurements from these next generation of galaxy surveys, accurate prior knowledge on the distribution of matter is required. This requirement is typically phrased in terms of knowledge of the {\it total matter power spectrum}, $P(k)$, which quantifies the amount of statistical power in a given Fourier mode of the matter overdensity field. For the next generation of surveys, the total matter power spectrum needs to be known to at least within $1\%$ accuracy up to $k=10\,h/$Mpc to achieve the desired accuracy in cosmological parameter constraints \citep{Huterer05,Laureijs09,Hearin12}.

In the past, it was sufficient to model the total matter power spectrum via analytical techniques \citep[e.g.][]{CAMB,Takahashi12} or using dark matter-only (DMO) simulations \citep[e.g.][]{coyote}. Recently, \citet{vanDaalen11} demonstrated that baryonic effects can have a significant impact on the distribution of matter, which needs to be incorporated into weak lensing analysis pipelines. The main effect to model is the suppression of power at scales of a few Mpc associated with gas ejected by Active Galactic Nuclei (AGN). Cosmological hydrodynamical simulations can provide these predictions, and results from several state-of-the-art simulations are available in the literature \citep{vanDaalen11,Illustris,Hellwing16,MBiipowerspec,IllustrisTNG}. Such simulations differ in the numerical methods and the implementation of baryonic (`sub-grid') processes, which can in turn result in varying predictions for the total matter power spectrum at small scales, where these physical processes are relevant.

Effective analytical models to account for the impact of baryons on the total distribution of matter have also been devised. Some of these consist of modifications of the `halo model' \citep{Seljak00,Semboloni13,Fedeli14,Mead15}, others of effective parameterisations of the transfer of power produced by the presence of baryons based on observational constraints and/or simulation results \citep{Mohammed14,Schneider15}. Several techniques to mitigate the presence of baryons have been proposed: marginalisation over the parameters of effective models \citep{Semboloni11} or over the principal components in linear combinations of observables that are most strongly affected by baryonic effects \citep{Eifler15,Kitching16}. The success of these techniques depends on the flexibility of the models to capture the true underlying matter distribution \citep{Mohammed17}.

In this work, we present results on the impact of baryons on the distribution of matter from the Horizon set of simulations, a state-of-the-art set of simulations with full implementation of baryonic physics~\citep{Dubois14,Dubois16}. The Horizon set comprises three simulations with the same volume and initial conditions. The main run includes all baryonic physics processes, a second run lacks AGN feedback and the third run is a DMO box for comparison. Horizon differs from other cosmological hydrodynamic simulations in several aspects. The numerical method implemented is based on the ``adaptive-mesh-refinement'' (AMR) technique and the only requirement on sub-grid parameters is such that the simulation matches the observed stellar mass-black hole mass relation and the black hole mass-velocity dispersion ($M_{\rm BH}-\sigma$) relation at $z=0$. The full physics run, Horizon-AGN, has been shown to be in good agreement with observations of the star formation history of the Universe, and colours and luminosity functions of galaxies across a wide range of redshifts, as shown by \citet{Kaviraj17}. These authors nevertheless identified an excess of low-mass red galaxies at low redshift, which was attributed to supernovae feedback being too inefficient in preventing the formation of these galaxies.

We quantify the impact of baryons and the role of AGN feedback on the distribution of matter across the range of scales and redshifts of interest to weak gravitational lensing surveys.  We compare our results to those from other groups and we test the applicability of the baryonic correction (BC) model of \citet{Schneider15} using our results. We make tables of the total matter power spectrum from the Horizon set publicly available\footnote{\url{https://www.horizon-simulation.org/catalogues.html}}.

This manuscript is organised as follows. Section \ref{sec:simulation} describes the set of cosmological simulations used in this work. In section \ref{sec:pispec4}, we describe the method for computing the matter power spectra. Section \ref{sec:bcm} describes the baryonic correction model of \citet{Schneider15}. We present our results in section \ref{sec:results}, followed by a discussion and conclusions in sections \ref{sec:discuss} and \ref{sec:conclude}, respectively. Unless otherwise noted, we adopt for this work a set of cosmological parameters consistent with those derived by the {\it Wilkinson Microwave Anisotropy Probe} team ({\it WMAP}7, \citealt{komatsuetal11}), in accordance with the set-up of the hydrodynamical simulations, described in detail in the following section.

\section{The Horizon simulation set}
\label{sec:simulation}

The Horizon set of simulations comprises three cosmological simulation boxes of $L=100 \, h^{-1}\rm\,Mpc$ on each side ran using the AMR code {\sc ramses}~\citep{teyssier02}: Horizon-AGN (with full baryonic physics implementation), Horizon-noAGN (solely lacking AGN feedback in comparison to Horizon-AGN) and Horizon-DM (a DMO run). The three simulations share the same initial conditions and cosmological parameters. For this work, this is crucial, as we are particularly interested in the comparison between the three simulations. All of the runs adopt the cosmological parameters obtained by {\it WMAP}7: a total matter density of $\Omega_{\rm m}=0.272$, a baryon density of $\Omega_{\rm b}=0.045$, a dark energy density of $\Omega_\Lambda=0.728$, an amplitude of the matter power spectrum determined by $\sigma_8=0.81$, a Hubble constant of $H_0=70.4$ km$/$s Mpc$^{-1}$, and the index of the primordial power spectrum given by $n_s=0.967$. 

There are $1024^3$ dark matter (DM) particles in each box. The dark matter mass resolution is $M_{\rm DM, res}=8.3\times 10^7 \, \rm M_\odot$ for the baryonic runs and $M_{\rm DM, res}=9.9\times 10^7 \, \rm M_\odot$ for Horizon-DM. This difference in the mass of the DM particles between the DMO run and the baryonic run is due to keeping $\Omega_m$ constant between simulations.

Details on the prescription for star formation, gas cooling and the refinement scheme are available in our previous work \citep{Dubois14}. For the purposes of this work, it suffices to remind the reader that stellar feedback is implemented in both baryonic runs of the Horizon suite and that this mode of feedback is not expected to affect the scales probed in this work. On the other hand, gas cooling is important for determining the distribution of matter at small scales. In the Horizon runs, it is implemented by means of hydrogen and helium cooling down to a temperature of $10^4\,{\rm K}$ including the contribution from metals \citep{sutherland&dopita93}. The metallicity of the gas is modelled as a passive variable, changing according to the injection of gas ejecta from stellar winds and supernovae explosions. In what follows, we focus here on describing the implementation of AGN feedback in detail, since our results and the comparison to other simulation suites are sensitive to this sub-grid model in particular.

Black holes are seeded in Horizon-AGN with a seed mass of $10^5$ M$_\odot$ whenever the gas cell density exceeds the hydrogen number density threshold $n > n_0$ where $n_0 = 0.1$ H$/$cm$^3$. Black holes are not allowed to form within $50$ kpc of an existing black hole \citep{duboisetal10}, and all black hole formation stops at $z=1.5$ \citep{Volonteri16}. 
  
Once formed, black holes accrete using a Bondi-Hoyle-Lyttleton accretion prescription. The rate of accretion is given by $\dot M_{\rm BH}=4\pi \alpha G^2 M_{\rm BH}^2 \bar \rho / (\bar c_s^2+\bar u^2) ^{3/2},$ where $M_{\rm BH}$ is the black hole mass, $\bar \rho$ is the local average gas density, $\bar c_s$ is the local average sound speed, $\bar u$ is the local average gas velocity relative to the black hole velocity, and $\alpha$ is a dimensionless boost factor. This boost factor allows us to compensate for our inability to capture the colder, denser regions of the interstellar medium due to lack of resolution. It is given by \citep{Booth09}
  \begin{equation}
    \alpha = \begin{cases} (n/n_0)^2, & \mbox{if } n>n_0 \\ 1, & \mbox{otherwise. } \end{cases}
\end{equation}

The effective accretion rate onto black holes is not allowed to exceed the Eddington accretion rate: $\dot M_{\rm Edd}=4\pi G M_{\rm BH}m_{\rm p} / (\epsilon_{\rm r} \sigma_{\rm T} c),$ where $\sigma_{\rm T}$ is the Thompson cross-section, $c$ is the speed of light, $m_{\rm p}$ is the proton mass, and $\epsilon_{\rm r}$ is the radiative efficiency, assumed to be equal to $\epsilon_{\rm r}=0.1$ for the \cite{shakura&sunyaev73} accretion onto a Schwarzschild black hole.

Thermal and kinetic feedback from AGN are implemented in Horizon-AGN as proposed by \citet{duboisetal12agnmodel}. At low accretion rates, feedback is in the form of bipolar outflows (``jets'') with wind velocities of $10^4$ km$/$s \citep{Omma04}, aligned with the spin of the black hole. At high accretion rates, AGN feedback is thermal, with energy deposited isotropically into a sphere of radius $2 \Delta x_{min}$  around the black hole. For a given Eddington ratio $\chi=\dot{M}_{\rm BH}/\dot{M}_{\rm Edd}$, the energy is deposited into the two modes as follows:
  \begin{equation}
    \dot{E}_{\rm AGN} = \begin{cases} 0.15\epsilon_{\rm r}\dot{M}_{\rm BH}c^2, & \mbox{if } \chi>0.01 \\ \epsilon_{\rm r}\dot{M}_{\rm BH}c^2, & \mbox{if } \chi\leq0.01 \end{cases}
\end{equation}
with a fixed radiative efficiency $\epsilon_r =0.1$. Due to the evolution of black hole accretion rates with cosmic time, the vast majority of AGN are in isotropic (``quasar'') mode at $z>2$, while most are in jet mode at lower redshift \citep{Beckmann17}. The minimum heating temperature adopted for Horizon-AGN is effectively null, allowing for continuous AGN feedback in the quasar mode. The Horizon-noAGN \citep{Peirani17} simulation lacks AGN feedback altogether, which allows us to isolate the impact of this particular mechanism on the total distribution of matter.

\section {Power spectra computation}
\label{sec:pispec4}

The distribution of matter is quantified through its power spectrum, $P(k)$. If the density field in the simulation at a given redshift is given by $\rho({\bf x},z)$, we can characterise the inhomogeneities in this field via $\delta({\bf x},z) = \rho({\bf x},z)/\bar{\rho}(z) - 1$, where $\bar{\rho}(z)$ is the mean density of the universe at a certain redshift. The Fourier transform of $\delta({\bf x},z)$ is labelled $\tilde{\delta}({\bf k},z)$. The statistical properties of these inhomogeneities are described via the power spectrum,
\begin{equation}
  \langle \tilde\delta({\bf k},z)\tilde\delta({\bf k}',z)\rangle = (2\pi)^3P(k)\delta_D^3({\bf k}-{\bf k}'), 
\end{equation}
with $\delta_D^3$, the Dirac delta function. As the power spectrum has units of volume, we also work in terms of the dimensionless quantity $\Delta^2$, which is related to the power spectrum by
\begin{equation}
\Delta^2(k) \equiv \frac{k^3}{2\pi^2}P(k).
\end{equation}
Nevertheless, we are most often interested in ratios between power spectra, which are insensitive to whether we are working with $P(k)$ or $\Delta^2(k)$.

The computation of the total matter power spectrum requires the mapping of each matter component onto a three-dimensional grid. In the case of DM, stars, and black holes, the mapping involves the application of a kernel to smoothly distribute the mass of each particle over neighbouring cells. In the case of the gas, the simulation outputs are given in terms of an AMR grid, with varying spatial resolution. To account for this, we convert the gas density field into a distribution of effective particles. This is done by looping over all cells in the AMR grid and placing a particle with the total mass of the cells in the centre-of-mass of the group. As a result, regions that are more refined will have a higher number of particles per unit volume. The calculation of the power spectrum described below does not explicitly correct for this mapping from grid into effective particles. However, to ensure no bias is introduced by performing this step, we also consider the case where we ignore the grid refinement above some cut-off scale. The result is that our fiducial method guarantees better than 1\% convergence of the total matter power spectrum over our desired range of wave-vectors when compared to the case when sub-structures are averaged at scales of $12\, h^{-1}\, \rm  kpc$ and above. 

For each matter component (effective gas particles, DM particles, star and black hole particles) we map their mass to a uniform grid 1024 cells across a side using a piecewise quadratic spline \citep{hockney&eastwood81}, whose Fourier transform is
  \begin{equation}
    W(k) = \left[\frac{{\rm sin}(\pi k / 2 k_{\rm N})}{\pi k / 2 k_{\rm N}}\right]^p,
  \end{equation}
  where $k_{\rm N}$ is the Nyquist wave number and we adopt $p=4$ to reduce the amount of aliasing and shot noise \citep{Lipatov02,Jing05,Cui08}. Each grid is Fourier transformed using the {\tt FOUR3M} routine presented in \citet{thacker&couchman06}, which is then convolved with a Green's function to minimise errors from the mass mapping \citep{hockney&eastwood81}, resulting in a Fourier grid for each component.
Additionally, by summing these Fourier grids we construct a total matter grid. The power spectrum of any individual matter component or of the total matter is then the mean of the squares of the corresponding grid values within fixed $k$ bins. We have verified that adopting a 2048$^3$ grid does not impact our results on the total matter power spectrum. For calculating cross correlations we first multiply the corresponding components' Fourier grids together, and then take the mean of the resulting grid values within fixed $k$ bins.

The estimation of the auto-power spectra is affected by the discreteness of the tracers. In other words, there is an additive component to the power spectrum given by the contribution of ``shot noise'',
\begin{equation}
  P_{\rm shot} = \frac{V}{N_{\rm eff}},
\end{equation}
where $V$ is the simulation volume. $N_{\rm eff}$ is the effective number of particles, which accounts for their difference in mass: $N_{\rm eff}=(\sum_i^N m_i)^2/(\sum_i^N m_i^2)$, and where $N$ is the number of particles and $m_i$, their individual masses \citep{Peebles93}. In this analysis, we present total matter power spectra after subtracting the shot noise component. We have verified, however, that this subtraction does not modify our results given that we are usually restricted to scales where this component is sub-dominant.

Finally, the outputs of the different Horizon simulation runs that we compare in this work can differ slightly in the value of the scale factor. We account for these differences by performing a linear re-scaling based on the predicted linear growth function for our adopted cosmology, $D(z)$. In linear theory, the power spectrum at a given $z$ can be obtained by re-scaling the $z=0$ power spectrum: $P(k,z)=D^2(z)P(k,0)$. There are limitations associated with this re-scaling, which can lead to residual differences in the matter power spectrum at large scales. For a detailed discussion of this effect, see Appendix \ref{app_growth}. Cosmic variance can also have an impact in our predictions, which is discussed in Section \ref{sec:cv}. Appendix \ref{app_res} presents several convergence tests of our results.

\section{Baryonic correction model}
\label{sec:bcm}

The BC model developed by \citet{Schneider15} was proposed to account for the impact of baryons on the total matter power spectrum by modifying the density field of dark-matter-only $N$-body simulations to mimic the effects of baryons from any underlying adopted feedback recipe. We summarise the BC model here and compare Horizon results to this model in the following section.

The main assumption behind the BC model is that haloes can be decomposed into four constituents: hot gas in hydrostatic equilibrium, ejected gas from feedback processes, stars from a central galaxy, and adiabatically relaxed dark matter. These four components alter the total distribution of matter, compared to that from dark matter-only simulations, by generating an excess at small scales due to efficient cooling of the gas leading to star formation, and a suppression at intermediate scales which depends on the mass fraction of gas ejected by the AGN and its corresponding ejection radius. The components of the model are constrained by a combination of low resolution hydro-dynamical simulations and observations. In particular, one wishes to have accurate models for the abundance fraction of each matter component, and its spatial profile. For example, \citet{Schneider15} adopt a parametrisation of the fraction of stars in a central galaxy proposed by \citet{Kravtsov14}, and a stellar profile following results from simulations of galaxy clusters by \citet{Mohammed14b}.

Putting together the different model components, and studying the change in the predictions for a wide range of parameter space, \citet{Schneider15} suggested that an effective parametrisation of the impact of baryons on the total matter power spectrum would require capturing the {\it amount and scale of suppression} driven by gas ejection and the {\it enhancement of the small-scale power spectrum due to the stellar component}. As a consequence, the authors proposed a simplified parametrisation of baryonic effects on the matter power spectrum in the form of the product of two functions which represent these two effects:
\begin{equation}
  F(k,z) \equiv \frac{P_{\rm BCM}}{P_{\rm DMO}} = G(k|M_c,\eta_b,z)S(k|k_s),
  \label{eq:bcm_fkz}
\end{equation}
where $P_{\rm BCM}$ is the total matter power spectrum (``BCM'' stands for the BC model) and $P_{\rm DMO}$ is the matter power spectrum for a dark-matter-only simulation with the same cosmology. $G$ is a function that captures the effect of AGN on the distribution of matter, through the ejection of gas, and $S$ represents the impact of star formation and baryonic cooling at small scales.

The suppression due to gas ejected by AGN is parametrised with the following function
\begin{equation}
 G(k|M_c,\eta_b,z)=\frac{B(z)}{1+[k/k_g(z)]^3}+[1-B(z)],
\end{equation}
where $B(z)$ parametrises the redshift dependence of the power suppression due to AGN feedback, with a characteristic redshift $z_c=2.3$,
\begin{equation}
  B(z) = B_0\left[ 1+\left(\frac{z}{z_c}\right)^{2.5} \right]^{-1},
\end{equation}
and an amplitude $B_0$ related to the mass of the galaxy clusters typically responsible for the suppression, $M_c$,
\begin{equation}
  B_0=0.105\log_{10}\left(\frac{M_c}{{\rm M_\odot}/h}\right)-1.27.
\end{equation}
The function $k_g(z)$ sets the typical scale of the gas ejection, 
\begin{equation}
  k_g(z)=\frac{0.7[1-B(z)]^4\eta_b^{-1.6}}{h/{\rm Mpc}},
\end{equation}
parametrised by $\eta_b$, a parameter which relates the virial radius of the cluster to the distance at which the gas is ejected.

The stellar profile enhances the total matter power spectrum below scales of $k_s=55\,h \, \rm Mpc^{-1}$ with a quadratic polynomial
\begin{equation}
S(k|k_s) = 1+(k/k_s)^2.
\end{equation}

The redshift dependence of the BC model has not been explored in detail. \citet{Schneider15} assumed that the model parameters were redshift independent in their original study, and discussed the predictions of the BC model in the range $z=0-2$ under this assumption. By comparing the BC model to the Horizon predictions for the impact of baryons on the total matter power spectrum at different redshifts, we specifically test the validity of this assumption in the following sections.

\section{Results}
\label{sec:results}

\subsection{The impact of AGN feedback}
\label{sec:results1}

AGN feedback can have different consequences on the distribution of matter at small scales. It can heat the gas around a halo, preventing it from cooling, being accreted, and forming stars, or it can directly expel gas from a halo. We isolate the impact of AGN feedback on large-scale structures by comparing the total matter power spectrum of the Horizon-AGN and the Horizon-noAGN simulation runs.

\begin{figure}
\includegraphics[width=0.47\textwidth]{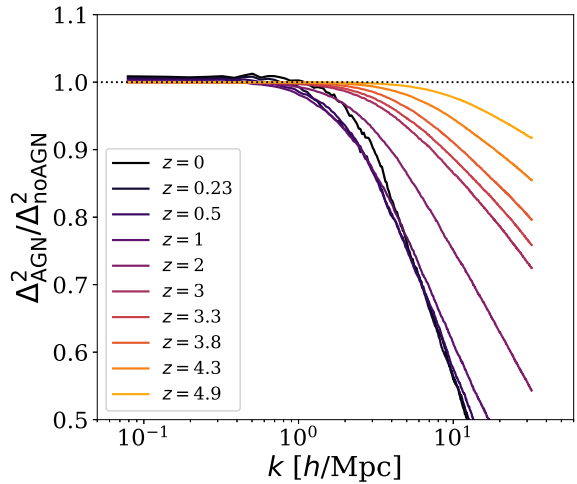}
\caption{The impact of AGN feedback in the total matter power spectrum from the Horizon set across cosmic time, $0\leq z<6$. Darker colours correspond to lower redshifts. At intermediate scales ($k \sim 2\,h/$Mpc), the AGN feedback suppression diminishes slightly from $z=1$ to $z=0$, an effect that we attribute to AGN feedback being insufficient to expel gas from the growing potential well of the most massive haloes.}
\label{fig:PScomp_AGN_noAGN}
\end{figure}

Figure \ref{fig:PScomp_AGN_noAGN} shows the ratio between the total matter power spectrum in Horizon-AGN and Horizon-noAGN at several redshifts, in the range from $z=4.9$ to $z=0$. (For a discussion on the numerical convergence of our results, see appendix \ref{app_res}.) As can be seen in Figure \ref{fig:PScomp_AGN_noAGN}, AGN feedback suppresses power at small scales ($k \geq 10\,h/$Mpc) as early as $z=4.9$, and the magnitude of this suppression increases towards low redshift. At intermediate scales ($k \sim 2\,h/$Mpc), the AGN feedback suppression diminishes slightly from $z=1$ to $z=0$.

Drawing on our previous analysis of the impact of AGN feedback on the quenching of star formation \citep{Beckmann17}, these results can be interpreted as follows. AGN feedback effectively regulates baryonic content at small scales, whether within smaller haloes or in the centre of larger ones, by heating the gas, redistributing it and preventing star formation. This process remains active throughout cosmic history. At redshifts $z \geq 1$, AGN in massive haloes drive large outflows and reduce inflows into their host galaxies, decreasing the power spectrum. In this redshift range, the suppression increases for any given $k$ as the black holes powering AGN continue to grow. Around $z=1$, several effects come into play that reduce the impact of AGN at a scale of a few $h/$Mpc. \citet{volonterietal16} showed that the biggest black holes at this redshift are accreting less efficiently. The combination of lower accretion rate and the transition to the jet regime at low redshift results in and overall decrease of feedback energy. As a consequence, gas accretion rates can increase at these redshifts. A potential interpretation of these results is that previously ejected gas can be re-accreted by haloes as they continue to grow. \citet{Beckmann17} have indeed shown an increase of inflow rates for massive galaxies at low redshifts in Horizon-AGN compared to Horizon-noAGN. As a result, gas would no longer be ejected to large scales and previously ejected gas could be re-accreted, so the suppression caused by AGN around $k \sim 2\,h/$Mpc decreases. This intepretation is supported by the non-monotonic trend in the fraction of gas within the virial radius of massive haloes, which decreases towards $z=1$, and grows thereon. Nevertheless, we note that it is also possible to obtain a decrease in gas fraction if the rate at which it is converted into stars increases. Massive haloes tend to dominate the matter power spectrum at $k \sim 2\,h/$Mpc, as shown by \citet{vanDaalen15}. The power spectrum remains suppressed at small scales as feedback continues to affect the centre of large haloes and the environment of smaller ones.

Figure \ref{fig:PScomp_AGN_noAGN} displays a large-scale excess of power below $1\%$ for several redshifts. This is intriguing because given the same initial conditions, finite volume effects are expected to cancel at large scales. In brief, the simulation outputs we compare for the curves shown in Figure \ref{fig:PScomp_AGN_noAGN} differ slightly in their scale factor ($<0.1\%$) and we have attempted to correct for the large-scale evolution of the power spectrum in the box by re-scaling it using the linear growth factor. However, even after this correction, our results are subject to $1\%$ biases of the matter power spectrum. These effects are sub-dominant compared to the impact of baryonic processes in which we are interested for this work. We have performed an extensive investigation of the source of this excess which we describe in detail in Appendix \ref{app_growth}.

\subsection{The impact of baryons}
\label{sec:impact}

\begin{figure}
\includegraphics[width=0.47\textwidth]{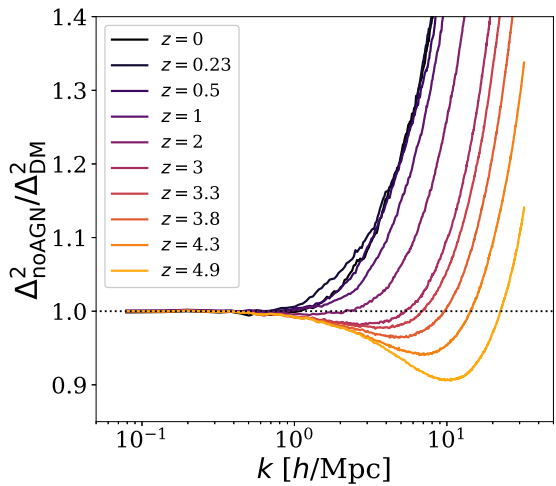}
\caption{The fractional impact of baryons on the total matter power spectrum when comparing the Horizon-noAGN run (lacking AGN feedback) to the Horizon-DM run.}
\label{fig:PScomp_noAGN_DM}
\end{figure}
\begin{figure*}
\includegraphics[width=0.95\textwidth]{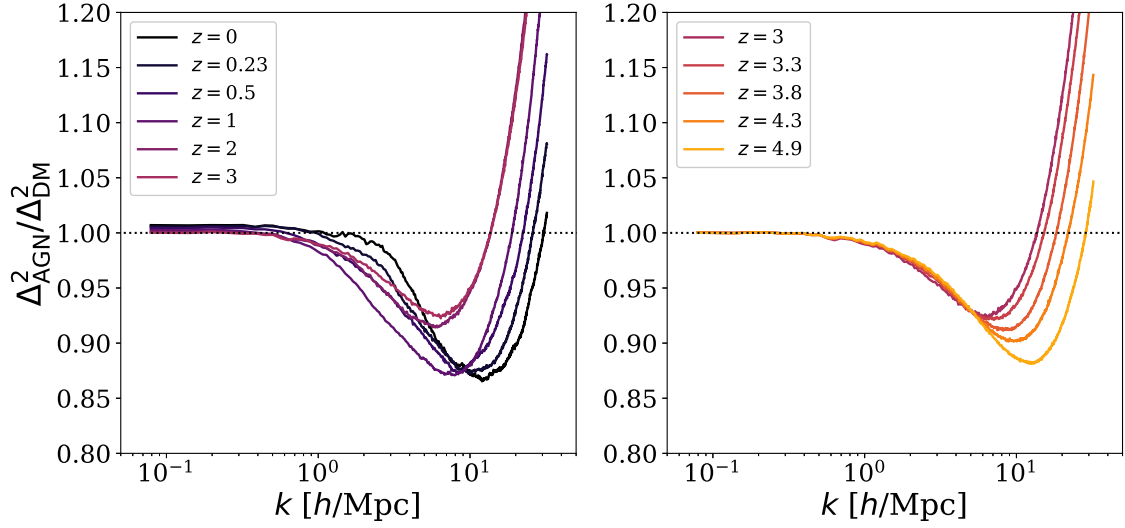}
\caption{The fractional impact of baryons on the total matter power spectrum when comparing the Horizon-AGN run (with AGN feedback) to the Horizon-DM run. Results are split in two panels for different redshift ranges: $z\leq 3$ (left panel) and $z\geq 3$ (right panel).}
\label{fig:PScomp_AGN_DM}
\end{figure*}

In this section, we compare the total matter power spectra obtained from the baryonic simulation runs to that from the Horizon-DM run. Our results are shown in Figure \ref{fig:PScomp_noAGN_DM} for Horizon-noAGN and Figure \ref{fig:PScomp_AGN_DM} for Horizon-AGN. The fractional impact of the effect of including baryons amply exceeds the $1\%$ requirement on the knowledge of this observable at $k=10\,h/$Mpc for future missions.

In Figure \ref{fig:PScomp_noAGN_DM}, we find that the impact of baryons on the total matter power spectrum at high redshift is to produce a suppression of power at scales above a few $h/$Mpc, accompanied by an enhancement at the smallest scales probed. This is not a consequence of AGN feedback, since this feedback mechanism is not present in Horizon-noAGN. The cause of the small scale enhancement is the additional cooling produced by the presence of baryons, which leads to an adiabatic contraction of the matter distribution at these scales \citep{Blumenthal86}. Compared to Horizon-DM, the Horizon-noAGN run also shows a suppression of power of approximately $10\%$ at scales of $k \simeq 10\,h/$Mpc at $z=4.9$. This is a consequence of the delayed collapse of DM haloes given the pressure contributed by the presence of baryons. Feedback processes do not have an impact on results at this redshift, neither from AGN nor from supernovae.\footnote{We have verified this by running two unrefined simulations, a pure dark matter and an adiabatic simulation with gas but no galaxy formation, of the same volume and initial conditions as the Horizon suite, and we have found this suppression to be present in the adiabatic run.} As redshift decreases, the suppression tends to be removed by the overall growth of structure.

Figure \ref{fig:PScomp_AGN_DM} shows the comparison between the Horizon-AGN run, including the impact of AGN feedback, and the Horizon-DM run, and is the main result of this work. The results at $z=4.9$ are very similar to those shown in the previous figure. This is a consequence of the small impact of AGN feedback at this redshift, which was evidenced in Figure \ref{fig:PScomp_AGN_noAGN} in the previous sub-section. 

The combined effect of the clustering of matter with the impact of AGN on its distribution leads to a non-monotonic redshift evolution of the ratios of power spectra between simulation runs. From $z=4.9$ to $z=3$, the effect of the AGN is not strong enough to compensate for adiabatic contraction. From $z=3$ to $z=1$, we find enhanced suppression due to the impact of AGN. From $z=1$ to $z=0$, the suppression is roughly constant but is shifted to smaller scales. As mentioned in the previous section, this behaviour is a consequence of haloes becoming too massive for AGN feedback to efficiently eject (or prevent the accretion of) material. As a result, clustering increases at intermediate scales of $k \sim 2\,h/$Mpc. In Section \ref{sec:cv}, we demonstrate that this specific behaviour is determined by the largest mass haloes formed in the simulation box.

\subsection{Dark matter response to baryons}

\begin{figure*}
\includegraphics[width=0.95\textwidth]{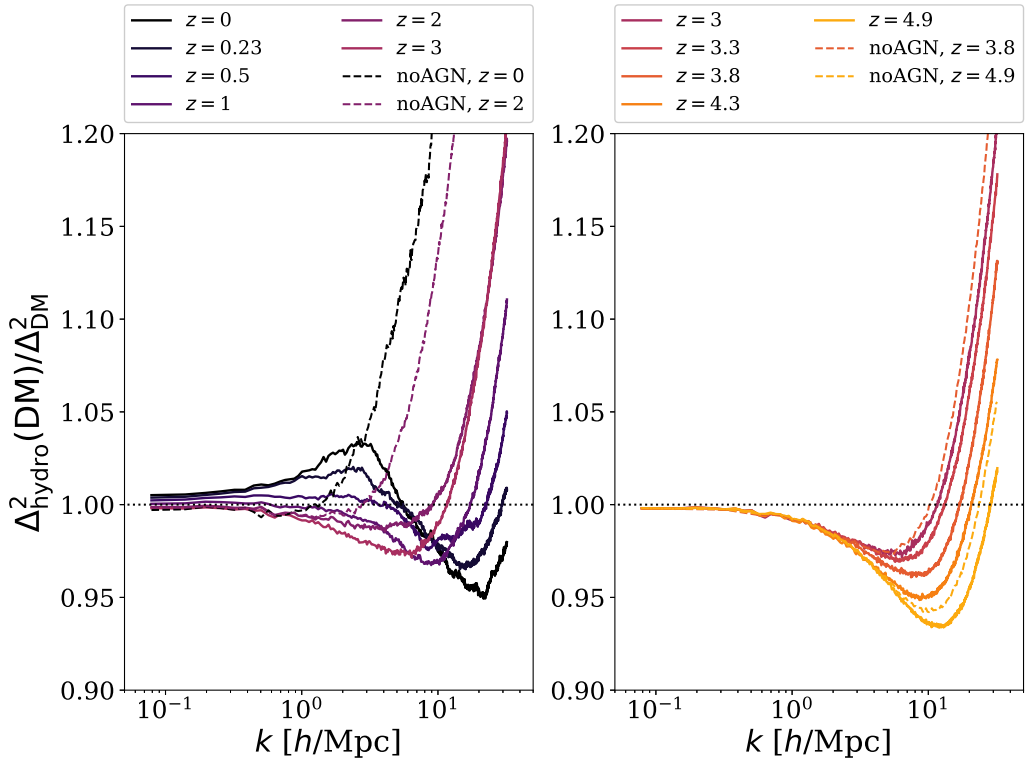}
\caption{The fractional power spectrum of the \emph{dark matter component} from the Horizon-AGN (solid) and the Horizon-noAGN (dashed) runs in the redshift range between $0\leq z< 5$. Only selected redshifts are shown for the Horizon-noAGN runs.}
\label{fig:PScomp_AGN_DM2}
\end{figure*}

\begin{figure*}
\includegraphics[width=0.47\textwidth]{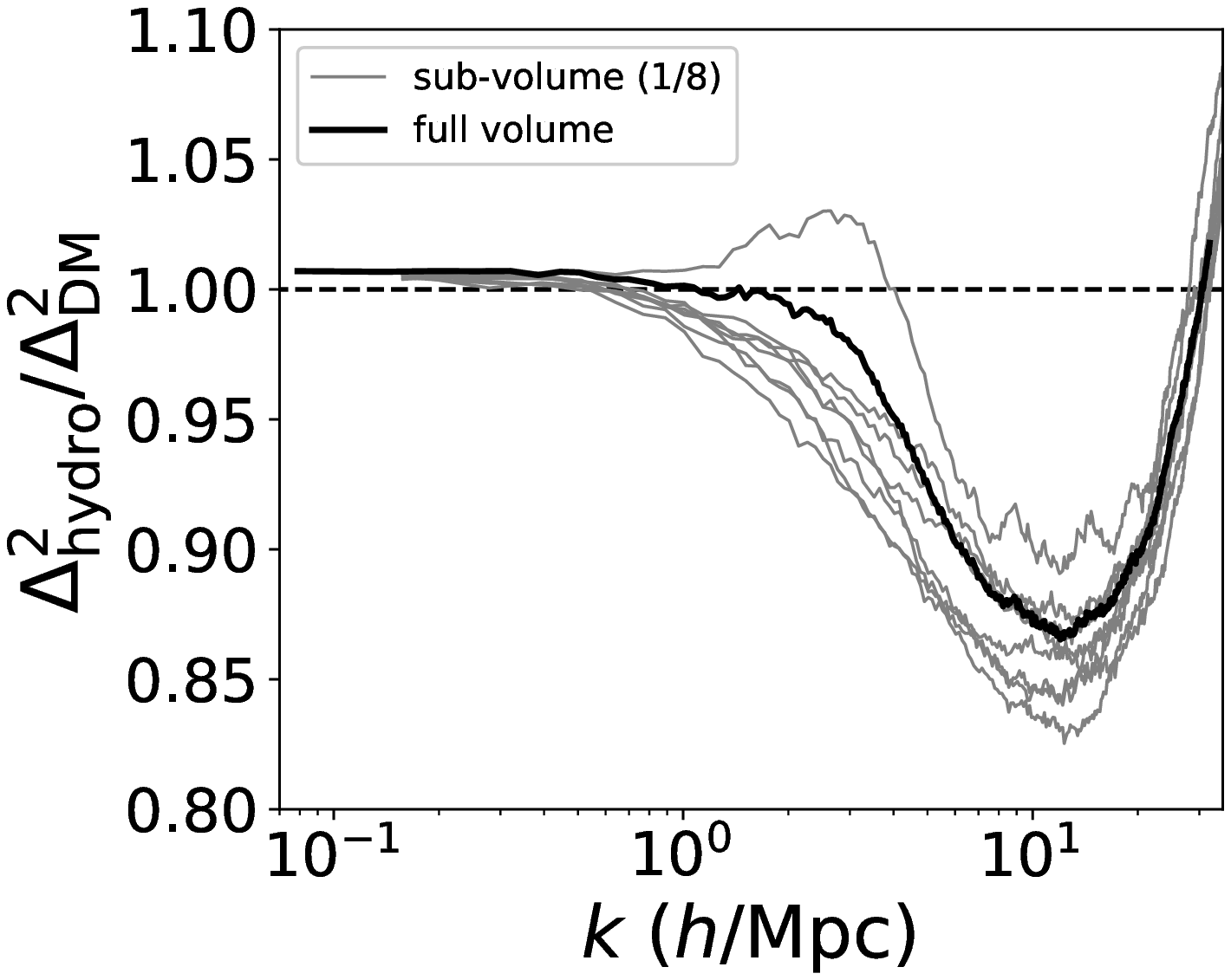}
\includegraphics[width=0.47\textwidth]{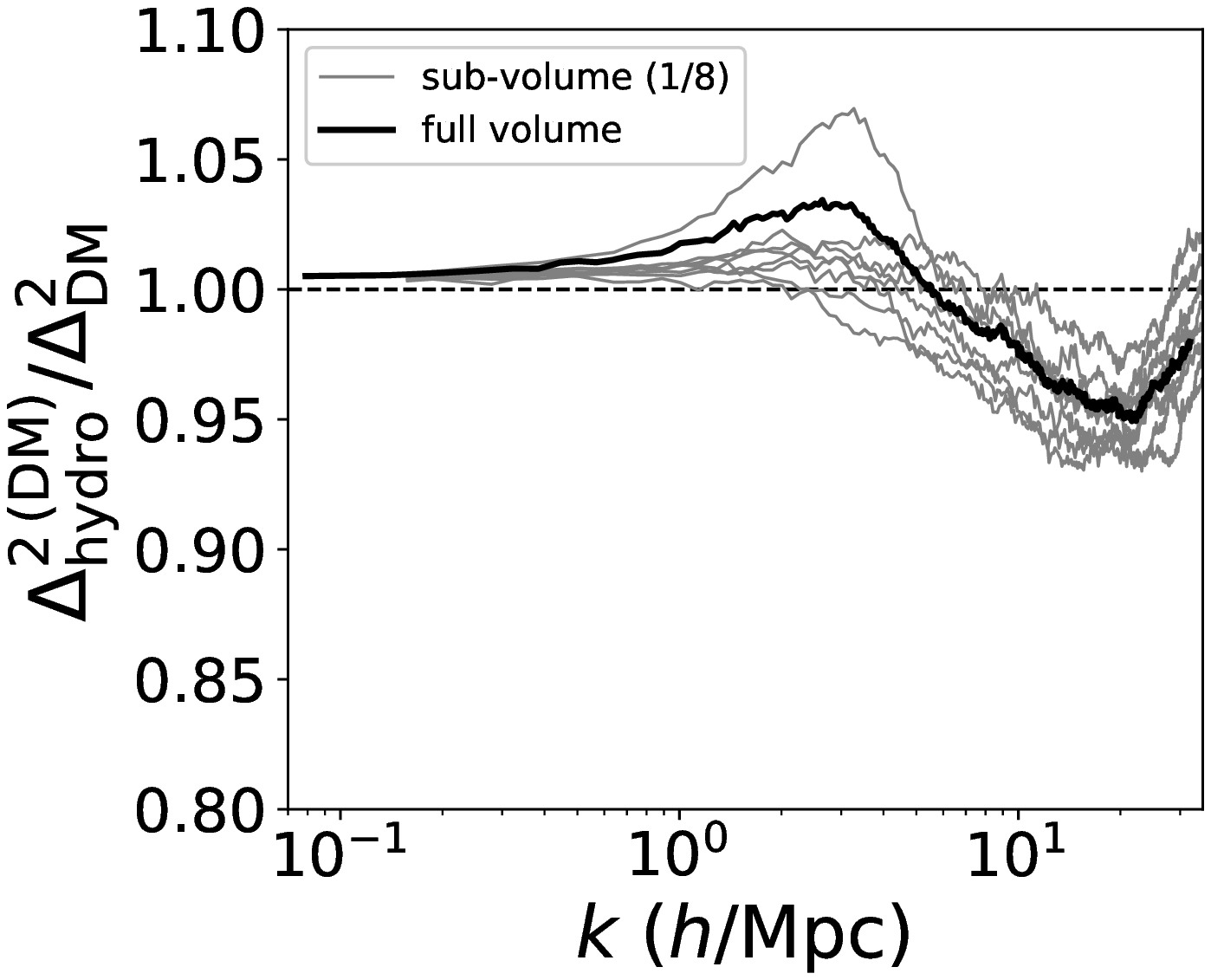}
\caption{The impact of cosmic variance on the total matter power spectrum (left) and on the power spectrum of the dark matter component (right) at $z=0$. The results for the full volume are shown in thick solid black; all other curves correspond to the $8$ sub-volumes of the box. The large-scale excess is discussed in Appendix \ref{app_growth}. We have not applied any correction for shot noise in this figure, given that it is sub-dominant at this redshift.}
\label{fig:subvol}
\end{figure*}

Do DMO simulations capture the dark matter component of the hydrodynamical simulations correctly? In other words, do baryons significantly affect the distribution of dark matter? We answer this question in Figure \ref{fig:PScomp_AGN_DM2}, where we show the ratio between the dark matter power spectrum in Horizon-AGN and Horizon-DM in the range $0\leq z< 5$. For comparison, we also show the ratio between Horizon-noAGN and Horizon-DM at selected redshifts only. 

At $z=4.9$ (right panel of Figure \ref{fig:PScomp_AGN_DM2}), the DM component in both Horizon-AGN and Horizon-noAGN runs has similar behaviour, as the AGN have not yet had time to cause a significant effect on the distribution of matter. Both curves show a $\sim 5\%$ suppression at high redshifts, which we attribute to the delayed collapse of haloes in the presence of baryons. As we go towards lower redshift and up to $z=2$, this suppression is reduced for both Horizon-AGN and Horizon-noAGN, overcome by the cooling of the baryons, which enhances the gravitational potential wells of haloes in the hydrodynamical simulations and leads to adiabatic contraction of the DM component. From this point on, the noAGN simulation continues to cluster towards $z=0$.

On the contrary, at $z=2$, we once again start to find a decrement of power at intermediate scales in the Horizon-AGN run. At lower redshift, the excess of power is transferred to smaller and smaller scales (haloes become more concentrated), while the suppression at scales of $\sim 10\,h/$Mpc is enhanced, and there is a compensation at scales of a few $h/$Mpc. As we discuss in Section \ref{sec:cv}, this specific scale-dependence is related to the limited cosmological volume of the simulation box, i.e., it is sensitive to cosmic variance.

At low redshift, the suppression of the power in the DM component can be attributed to the DM following the redistribution of gas as a consequence of AGN feedback. We have verified that indeed there is a strong correlation between the gas and DM fields of the Horizon-AGN simulation at low redshift which shows a similar suppression. However, in this case, the effect is a consequence of delayed cooling of the baryons. Below $z=2$, AGN are predominantly in the jet regime \citep{Volonteri16} and this mode deposits energy into a bipolar outflow that prevents hot gas from cooling, but drives only modest outflows.

\citet{Peirani17} performed an analysis of the impact of AGN feedback on the density profile of DM haloes in the Horizon suite which is consistent our results. Using a cross-matched sample of DM haloes, they found that the haloes in Horizon-AGN are more steep that their DMO counterparts at $z=5$ and $z=0$ and that the evolution is non-monotonic, with haloes at $z=1.6$ being more shallow. The authors suggested this is a consequence of successive phases of contraction and expansion, with a ``cusp re-generation'' happening at low redshift as a consequence of dwindling AGN activity.

\subsection{Impact of cosmic variance}
\label{sec:cv}

In Section \ref{sec:results1}, we asserted that the availability of baryonic and DMO runs with the same initial conditions allowed us to neglect finite volume effects at small wave-number values. This does not guarantee, however, that we are free from the impact of cosmic variance in our results. We explore the consequences of the limited volume of the simulation box by dividing the Horizon boxes in $8$ sub-volumes and obtaining predictions of the impact of baryons on the total matter power spectrum from these sub-volumes. The results are shown in the left panel of Figure \ref{fig:subvol}, where the black thick solid line corresponds to the full volume results, and all other lines correspond to different sub-volumes, all at $z=0$. 

The dispersion of the grey curves gives us an estimation of the impact of cosmic variance on our results. In particular, we note that one of the sub-volumes displays an excess of power at $k\sim 3\,h/$Mpc in the total matter distribution with respect to the DM distribution in Horizon-DM. The comparison to the black solid curve suggests that our results from the full volume box are dominated by this particular sub-volume. The right panel of Figure \ref{fig:subvol} presents similar results for the power spectrum of the DM distribution alone. Most curves show some level of excess above zero at $k\sim 3\,h/$Mpc, but the full volume results once more are dominated by one of the sub-volumes. The sub-volume with the largest excess at these scales is the only one to host haloes with masses above $10^{14.5} {\rm M}_\odot$. These two haloes have masses of approximate $10^{14.8} {\rm M}_\odot$ each. Since the shape of the DM power spectrum changes from $z=1$ to $z=0$, it is possible that the assembly of these massive structures has a role to play in this result.

Intriguingly, the fact that the IllustrisTNG300 simulation, with a larger box size of $300 \, \rm Mpc$, also displays this pattern (Figure \ref{fig:compare_TNG_dmback}) suggests that this is robust to increasing the simulation volume. \citet{vanDaalen15} investigated the contribution of haloes of different masses to the total matter power spectrum in a set of simulations of different volumes, and pointed out that scales of $k\sim 3\,h/$Mpc are typically dominated by the most massive ones, which is in line with our conclusions. Overall, our results suggest that the accuracy of predictions for the impact of baryons on the matter power spectrum would improve by running larger volume hydrodynamic simulations with multiple realisations of the initial conditions.

\subsection{Comparison to other hydrodynamical simulations}
\label{sec:comparesims}

\begin{figure}
\includegraphics[width=0.47\textwidth]{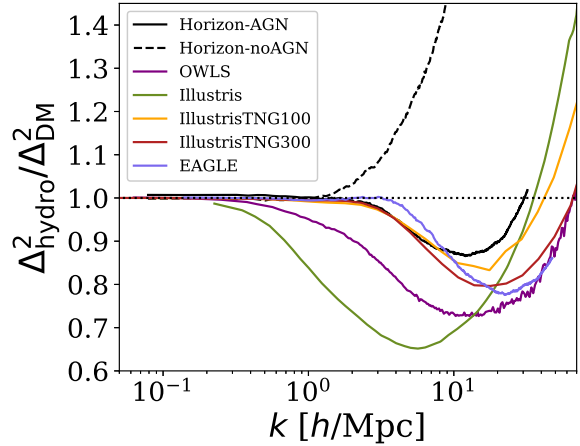}
\caption{The impact of baryons on the total matter power spectrum ($\Delta^2_{\rm hydro}/\Delta^2_{\rm DMO}$) in Horizon-AGN (solid black) and Horizon-noAGN (dashed black) compared to the results of other cosmological simulations at $z=0$.}
\label{fig:compare}
\end{figure}

Several other groups have quantified the impact of baryons on the matter power spectrum from their numerical simulations. Their simulations vary in the numerical technique implemented, volume, resolution and sub-grid recipes adopted for baryonic physics processes. In this section, we discuss how their results compare to Horizon-AGN. Figure \ref{fig:compare} shows a comparison of the fractional impact of baryonic processes on the total matter power spectrum from different simulations at $z=0$: the OverWhelmingly Large Simulations \citep[][we refer here to the `AGN' run of OWLS which adopts a {\it WMAP}7 cosmology]{vanDaalen11}, the EAGLE simulation \citep[][]{Hellwing16}, Illustris \citep{Illustris} and IllustrisTNG \citep{IllustrisTNG}. The result from Horizon-noAGN is also shown for reference, in which case there is an enhancement of power due to efficient cooling of the gas, rather than a suppression of power. For reference, the simulation volumes are as follows: OWLS and Horizon are $100$ Mpc$/h$ on each side; EAGLE, $100$ Mpc on a side; Illustris, $75$ Mpc$/h$ on a side and the IllustrisTNG runs are $100$ and $300$ Mpc on a side for ``TNG100'' and ``TNG300'', respectively.

While the qualitative behaviour of all simulations is similar, with a suppression of power due to the effect of AGN feedback on the gas at $k \sim 10\,h/$Mpc, the exact scale and strength of the suppression differs between them. Illustris shows the largest amount of suppression, reaching over $30\%$ at scales of $k\sim 5\,h/$Mpc. This simulation is calibrated to match the overall observed star formation history of the Universe, but despite this calibration, their radio mode of AGN feedback is known to be too aggressive, resulting in lower than observed gas fractions inside of massive haloes \citep{Haider16}.

The OWLS `AGN' run used by \citet{vanDaalen11} was calibrated to match the $M-\sigma$ relation \citep{Booth09,Schaye10}, similarly to Horizon-AGN, but differs in other sub-grid recipes (e.g., stellar initial mass function, stellar feedback prescription, black hole seeding, and thermal quasar AGN feedback for all accretion rates) and the numerical method implemented (smoothed-particle-hydrodynamics). \citet{McCarthy10} have shown that this OWLS run reproduces the fraction of gas in massive haloes and a further exploration, varying some of the sub-grid parameter models for the AGN feedback implementation, was performed by \citet{McCarthy11} and \citet{LeBrun14}. This is further discussed in Section \ref{sec:discuss}. At $z=0$, OWLS predict significantly more suppression than Horizon-AGN, exceeding $20\%$ at $k \sim 10\,h/$Mpc. The impact of baryons in the case of OWLS is not as strong as in the Illustris simulation. This model has been widely used in the literature for cosmic shear data analysis \citep{Mead15}, including recent cosmic shear survey results \citep[][]{Harnois15,Joudaki17,Krause17}, and also for forecasting the performance of future surveys \citep{Semboloni11,Semboloni13,Eifler15}.

The EAGLE simulation \citep{Schaye15} is a smoothed-particle-hydrodynamics simulation with similar volume to Horizon-AGN and full baryonic physics implementation. In this case, the simulation was calibrated to match the relation between stellar mass and halo mass, the present-day stellar mass function of galaxies and galaxy sizes. EAGLE predicts that the impact of baryons on the matter power spectrum is predominant at scales smaller than in Horizon-AGN, Illustris or OWLS. The difference in the preferred scale of suppression is particularly relevant to cosmic shear surveys adopting a cut on small scales in their analysis \citep{Krause17} instead of a marginalisation strategy \citep{Joudaki17}.

\begin{figure*}
\includegraphics[width=0.95\textwidth]{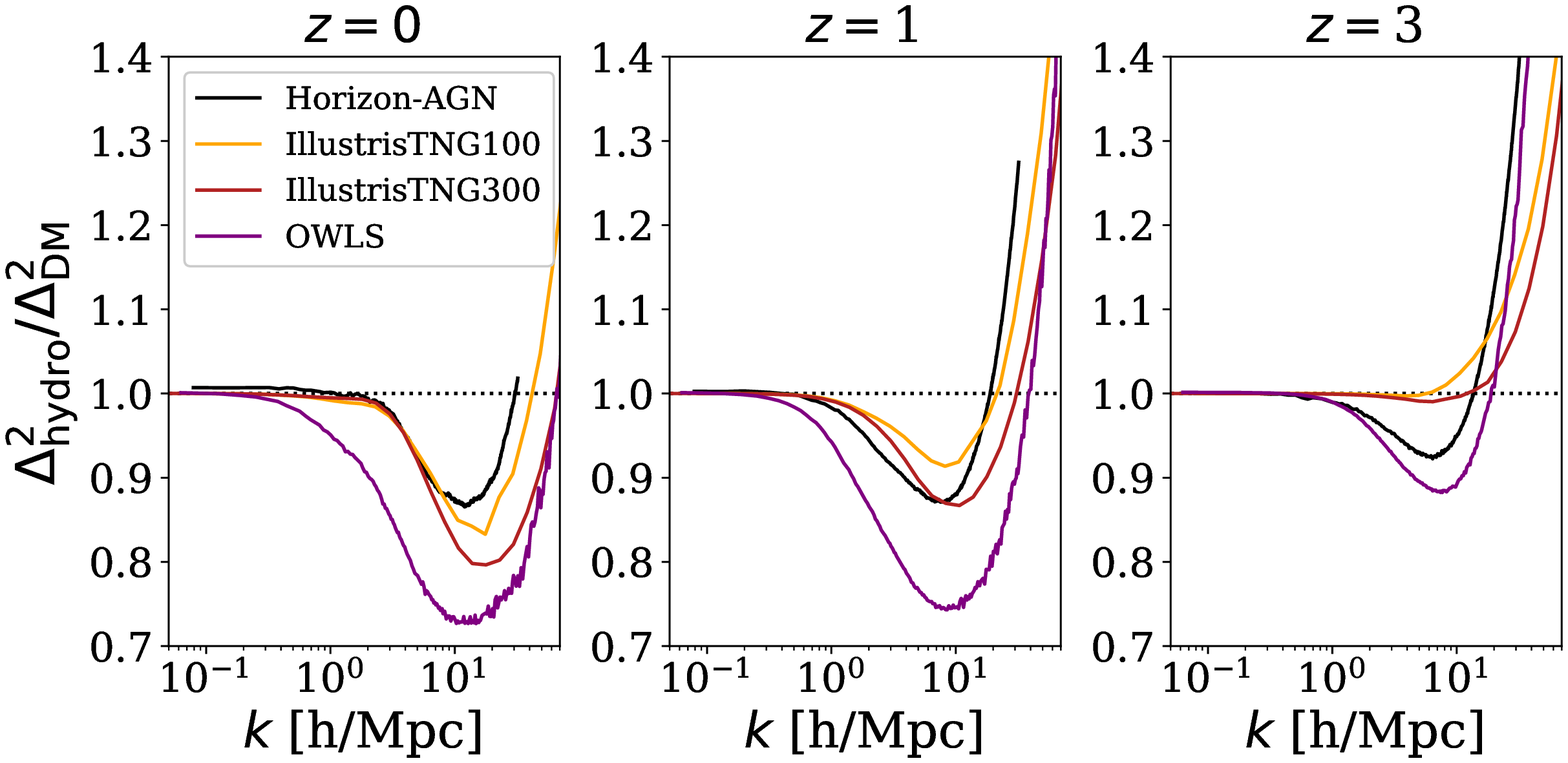}
\caption{The impact of baryons on the power spectrum at redshifts $z=0$ (left), $z=1$ (middle) and $z=3$ (right) for Horizon-AGN, OWLS and IllustrisTNG.}
\label{fig:compare_z}
\includegraphics[width=0.8\textwidth]{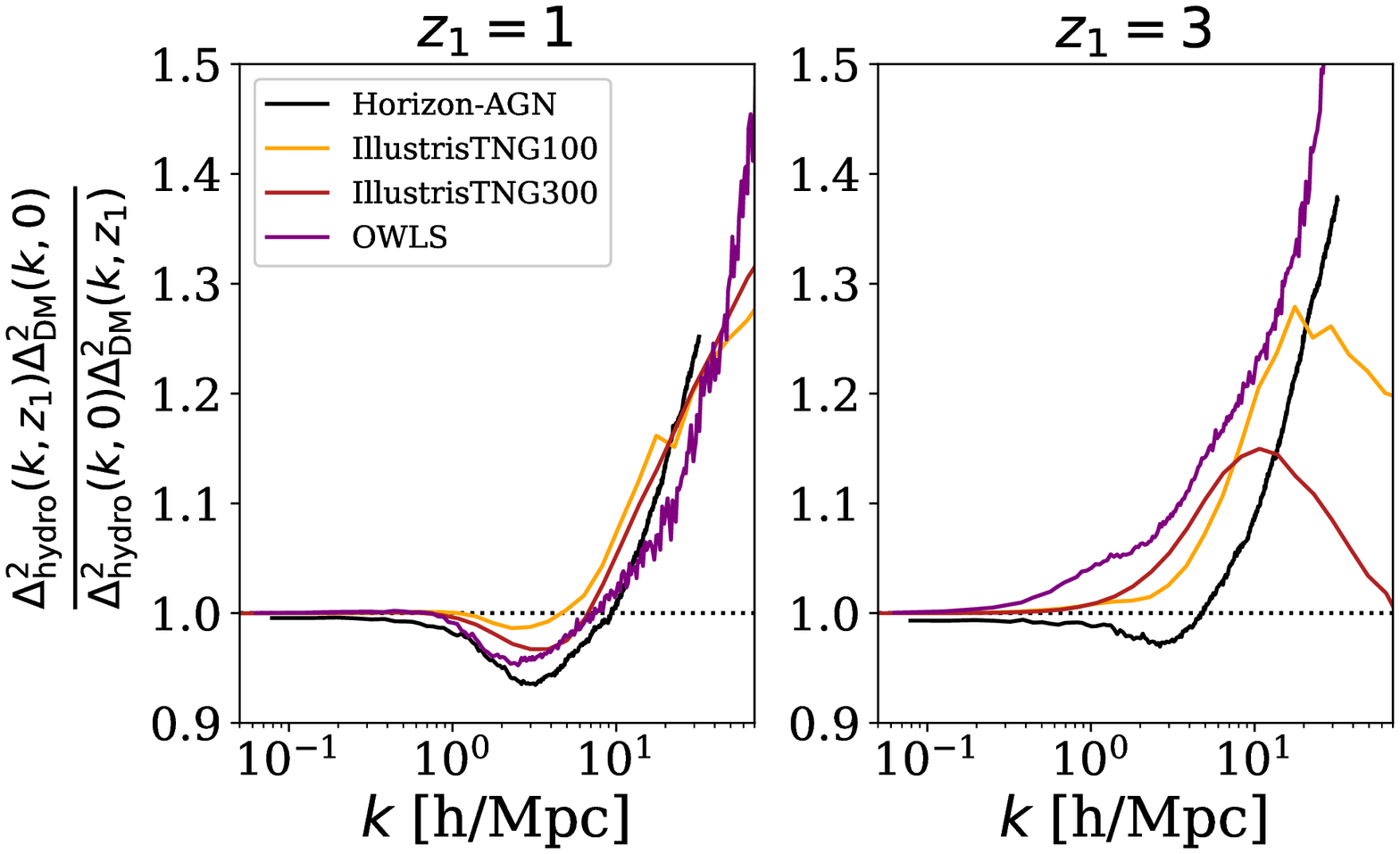}
\caption{The impact of baryons on the power spectrum at $z=0$ compared to $z_1={1,3}$. The $y$-axis can be interpreted as the ratio between effective scale-dependent growth functions: $\mathcal{D}^2(k,z) \equiv \Delta^2(k,z=0)/\Delta^2(k,z_1)$. It can also be understood as the relative redshift evolution of the impact of baryons on the power spectrum. The left panel corresponds to the growth from $z_1=1$ to $z=0$ and the right panel, from $z_1=3$ to $z=0$. Both panels compare different curves for Horizon-AGN, OWLS and IllustrisTNG. The evolution of the impact of baryons is similar between the simulation in the range from $z_1=1$ to $z=0$ but starts to differ at higher redshift.}
\label{fig:compare_z_growth}
\end{figure*}

\begin{figure}
\includegraphics[width=0.47\textwidth]{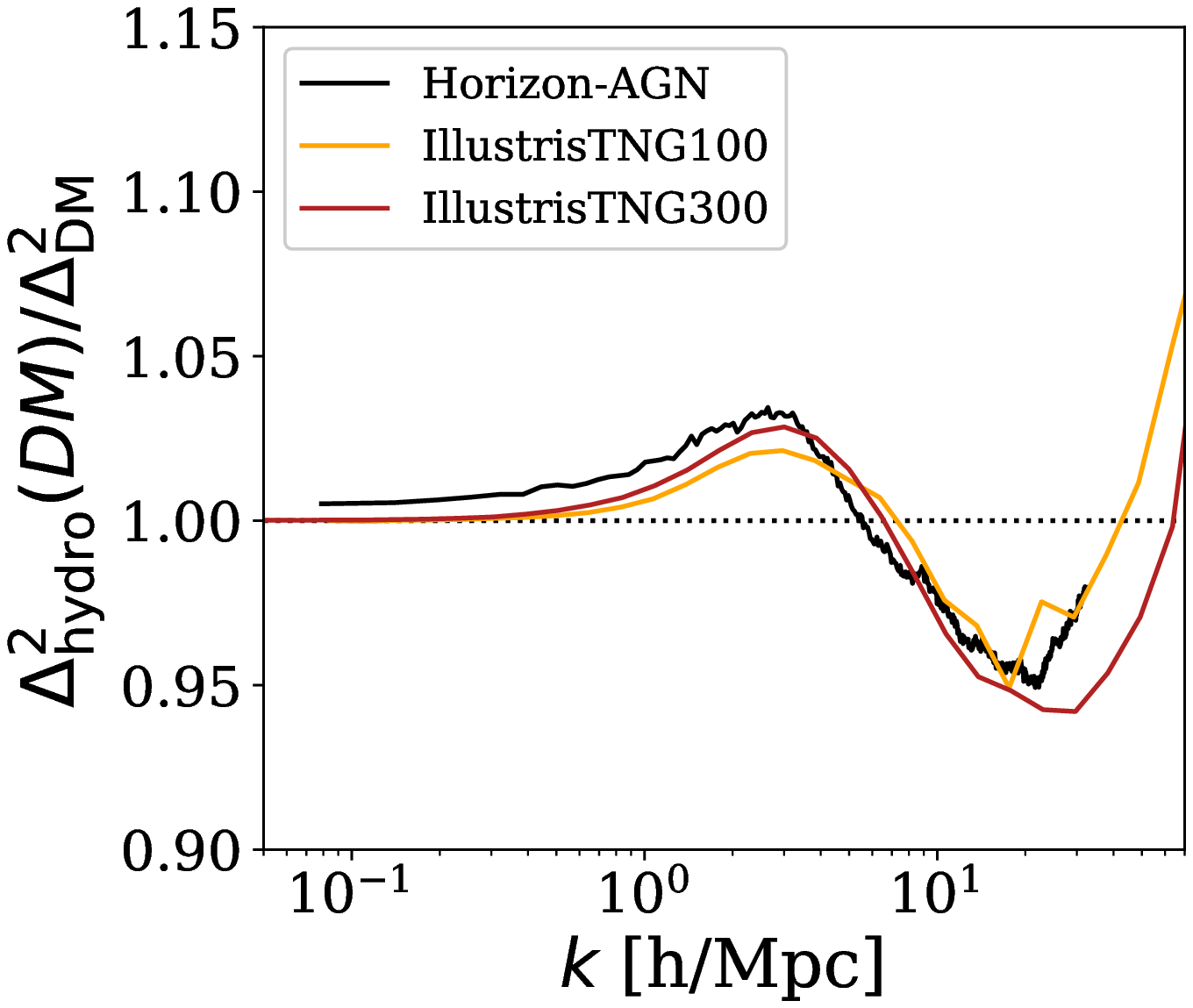}
\caption{A comparison of the sensitivity of the DM component power spectrum to baryonic effects in IllustrisTNG (orange and red for TNG100 and TNG300, respectively) and in Horizon-AGN (black).}
\label{fig:compare_TNG_dmback}
\end{figure}

\citet{IllustrisTNG} recently presented an analysis of the impact of baryons on the clustering of galaxies and matter in the IllustrisTNG simulations. IllustrisTNG is a set of cosmological simulation boxes with different volumes and physics implementations, and we are interested here in the comparison to the baryonic and DMO runs. IllustrisTNG implements an updated AGN feedback recipe compared to the previous Illustris runs \citep{Weinberger17}, among other changes (including SN feedback modelling). Their new AGN sub-grid model includes a different approach of radio kinetic feedback mode compared to Illustris, which one is very similar to the one implemented in Horizon-AGN~\citep{duboisetal12agnmodel} but with an isotropic momentum/energy injection (IllustrisTNG) instead of being jet-like shaped (Horizon-AGN). In Figure \ref{fig:compare}, we show the impact of baryons on the total matter power spectrum from IllustrisTNG100 and IllustrisTNG300 at $z=0$ as obtained by \citet{IllustrisTNG}. The new IllustrisTNG runs show significantly lower impact of baryons on the distribution of matter, with a reduction of the overall amplitude of the effect and a restriction to smaller scales compared to Illustris. The IllustrisTNG300 results are similar to those obtained by EAGLE, despite different numerical methods and sub-grid physics implementations. The discrepancy between IllustrisTNG100 and IllustrisTNG300 is attributed to differences in resolution and box size.

Compared to Horizon-AGN, the IllustrisTNG simulations present enhanced suppression of the total matter power spectrum and a displacement of the peak of the suppression towards small scales at $z=0$. \citet{IllustrisTNG} and \citet{vanDaalen11} also presented results at higher redshifts for IllustrisTNG and OWLS, respectively, which allows us to compare the redshift evolution across these simulations in Figure \ref{fig:compare_z}. Compared to Horizon-AGN, the redshift evolution is much more dramatic in OWLS and IllustrisTNG from $z=3$ to $z=0$. The Horizon-AGN total matter power spectrum already shows signs of suppression at $z=3$ due to AGN feedback (see discussion in Section \ref{sec:impact}), while IllustrisTNG only shows signs of gas cooling and adiabatic contraction at this redshift. IllustrisTNG undergoes a rapid redshift evolution towards $z=0$, overtaking the suppression found in Horizon-AGN. In Figure \ref{fig:compare_z}, we have limited our results to the redshift range of interest to weak lensing surveys and to the range limited by the convergence timescales of galaxy stellar populations. A comparison between OWLS and Horizon-AGN at higher redshifts yields a similar suppression at $z\simeq 3.8$, while the two simulations start deviating at even higher redshifts, with the suppression in OWLS becoming smaller while it increases in Horizon-AGN.

Figure \ref{fig:compare_z_growth} shows a slightly different rendering of the results presented in Figure \ref{fig:compare_z}. The two panels of Figure \ref{fig:compare_z_growth} show ratios of the impact of baryons between two redshifts: $z=0$ and $z_1$. The vertical axis represents an effective scale-dependent growth in the hydrodynamic simulation compared to the DMO case, defined as $\mathcal{D}^2(k,z_1) \equiv \Delta^2(k,z=0)/\Delta^2(k,z_1)$. The panels compare $z_1=1$ (left) and $z_1=3$ (right) with $z=0$. The vertical axis can also be interpreted as the relative redshift evolution of the impact of baryons on the power spectrum. From $z=1$ to $z=0$, this is qualitatively similar across simulations. Discrepancies arise when comparing $z=3$ and $z=0$, in which case both OWLS and IllustrisTNG show stronger redshift evolution than Horizon. In the case of IllustrisTNG, the downturn of the curves at large $k$ reflects a lack of suppression due to the presence of baryons at $z=3$. Weak lensing surveys which aim to constrain the redshift evolution of the parameters of the equation of state of dark energy should accommodate flexible models of the redshift evolution of this effect to avoid potential biases.

Finally, we have compared the impact of baryons in the DM distribution between Horizon-AGN and IllustrisTNG at $z=0$ in Figure \ref{fig:compare_TNG_dmback}. Despite the differences evidenced for the suppression of the total matter power spectrum, the impact of baryonic effects on the DM distribution is similar in Horizon-AGN, IllustrisTNG and (although not shown here), EAGLE \citep{Hellwing16}. All of these simulations feature enhanced clustering of the DM at very small scales (cuspier haloes due to adiabatic contraction), a suppression at intermediate scales and an enhancement at large scales. Nevertheless, \citet{IllustrisTNG} has shown (their Figure 8) that the Illustris simulation produces a widely different prediction in this case, with a suppression at scales of a few $h/$Mpc. It is possible that the strong radio mode of AGN feedback used in Illustris is dominating the scales of the problem. Similar results are found by \citet{vanDaalen11} in their Figure 6 for OWLS, where the AGN feedback is also much stronger than that in Horizon-AGN, EAGLE and IllustrisTNG.

\subsection{Effective modelling}
\label{sec:effmod}

\begin{figure*}
\includegraphics[width=0.47\textwidth]{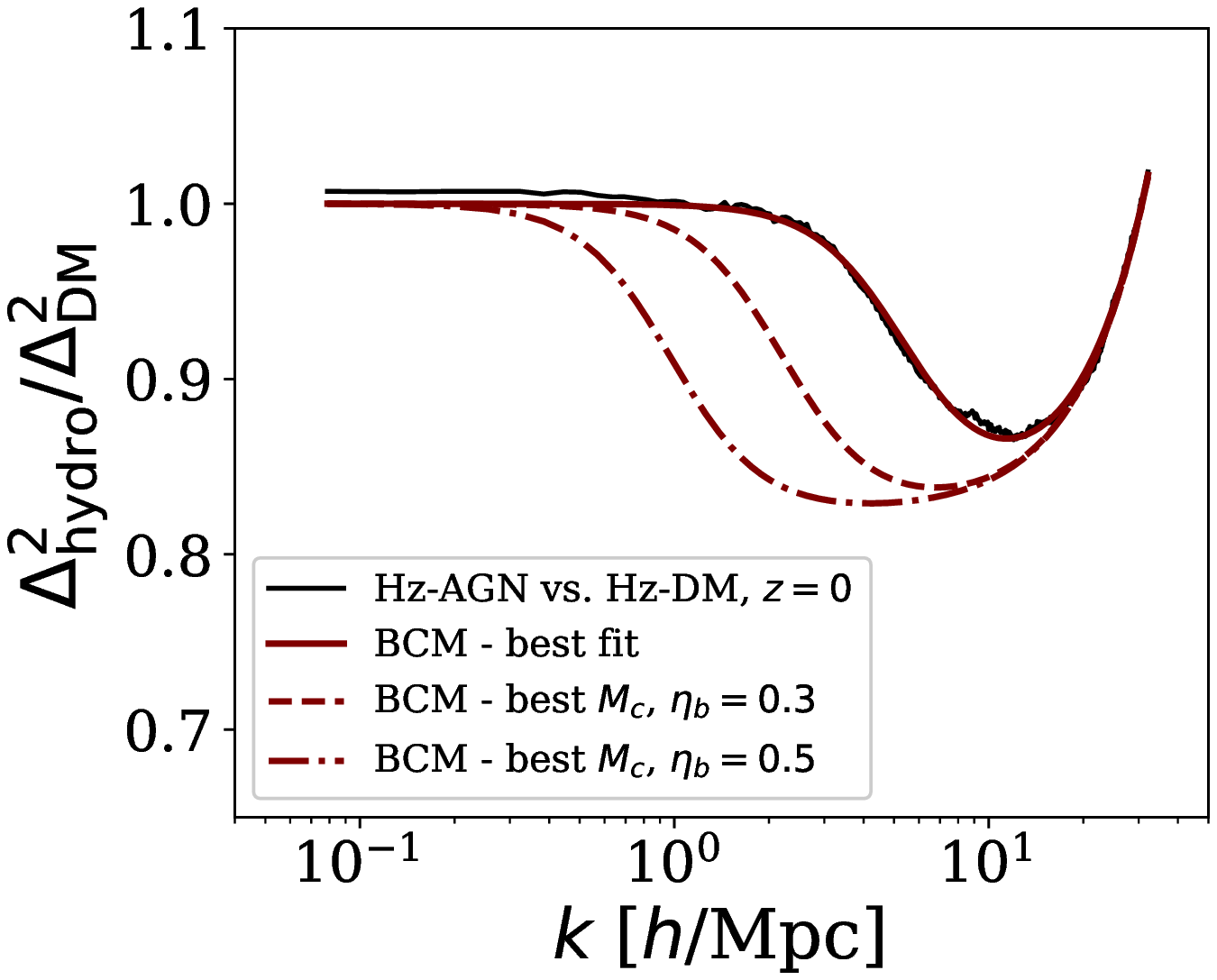}
\includegraphics[width=0.47\textwidth]{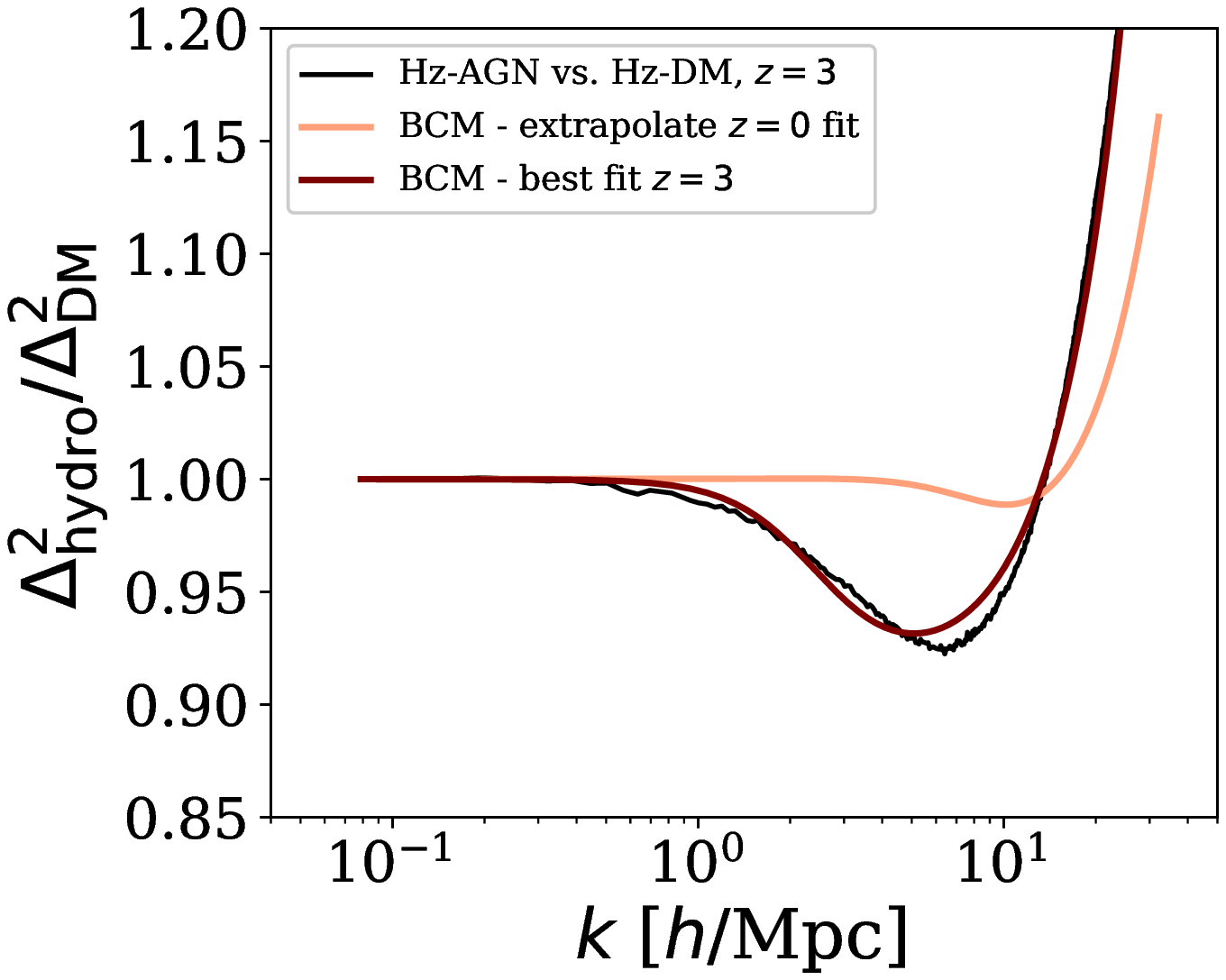}
\caption{The impact of baryons on the total matter power spectrum ($\Delta^2_{\rm hydro}/\Delta^2_{\rm DM}$) in Horizon-AGN and a comparison to the BC model prediction at redshifts $z=0$ (left panel) and $z=3$ (right panel). The black curves show the Horizon-AGN results. The red curves correspond to the best fit BC model, freeing the parameters $M_c$, $\eta_b$ and $k_s$ at each redshift. The left panel also shows the impact of changing the $\eta_b$ parameter to $0.3$ (dashed) and $0.5$ (dot-dashed), values which are better motivated by observations, for comparison. The $z=0$ fit is also extrapolated to $z=3$ in the right panel and compared to the Horizon results (orange). The fiducial BC model redshift evolution together with the $z=0$ best fit parameters does not match the BC model predictions.}
\label{fig:bcmcomp1}
\end{figure*}
\begin{figure*}
\includegraphics[width=0.9\textwidth]{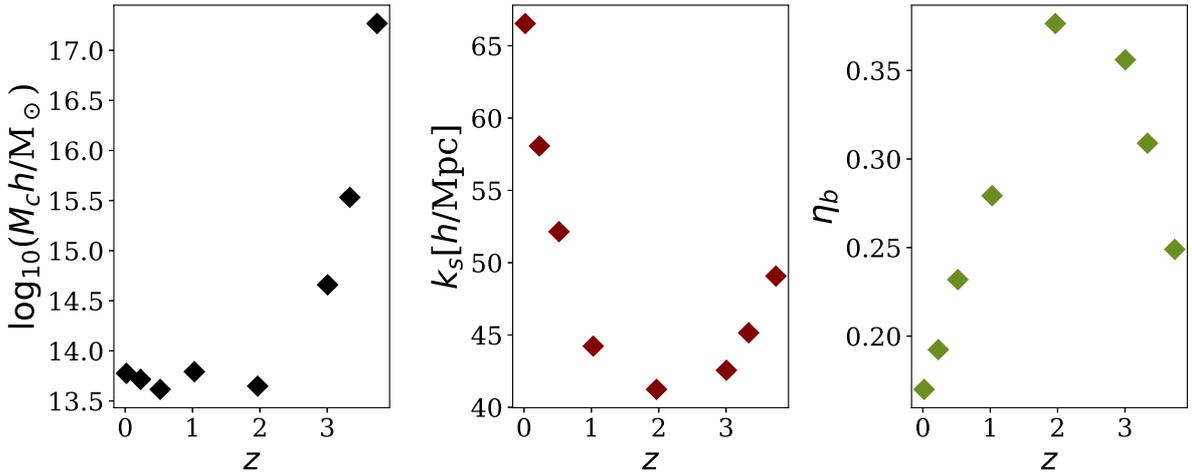}
\caption{The best fit BC model parameters in each simulation snapshot in the range $0<z<4$. The left panel shows preferred values of $M_c$; the middle panel, for $k_s$; and the right panel, for $\eta_b$. The fiducial BC model assumption is to keep these parameters constant as a function of redshift.}
\label{fig:bcmcomp2}
\end{figure*}
\begin{figure}
\includegraphics[width=0.47\textwidth]{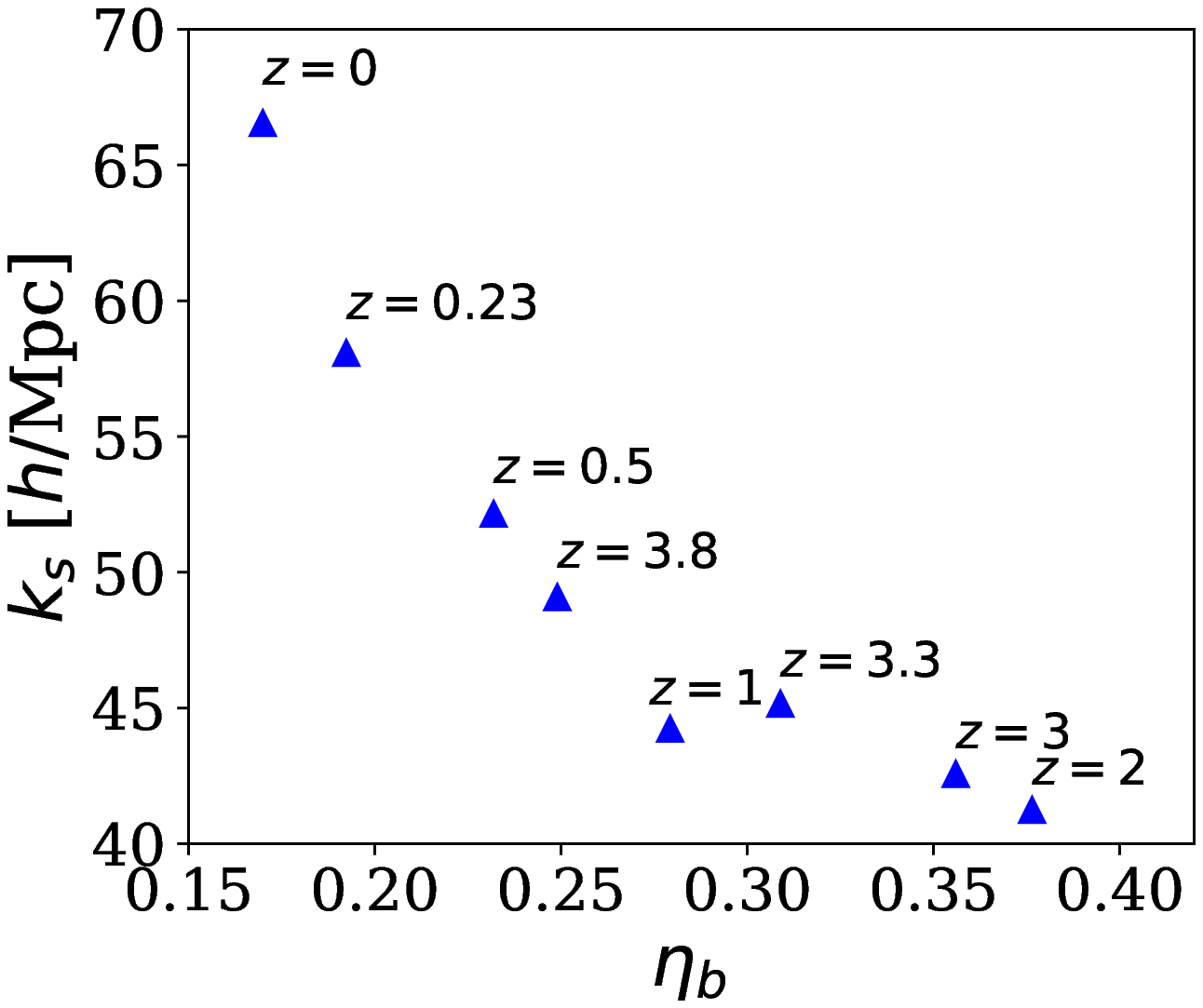}
\caption{$k_s$ and $\eta_b$ preferred values for the BC model at each redshift (indicated by the label on each point). We find an anti-correlation between the two parameters.}
\label{fig:anticorr}
\end{figure}

In this section, we discuss the application of the BC model introduced in section \ref{sec:bcm} to our simulation results. In \citet{Schneider15}, the BC model parameters were estimated from a combination of observational constraints and results from numerical simulations. These parameters are: $M_c$, which drives the amount of suppression in the power spectrum, $\eta_b$, the parameter that governs the scale of the suppression of power, and $k_s$, the wave-number associated to the slope of the stellar density profile of the central galaxy. Figure \ref{fig:bcmcomp1} shows a comparison between the Horizon-AGN prediction of the fractional impact of baryons on the matter power spectrum (solid black curves) at $z=0$ (left panel) and $z=3$ (right panel) compared to the best fit BC model (in red). 

The Horizon-AGN results are not well described by that fiducial choice of BC model parameters at $z=0$. However, this should not come as a surprise since the BC model assumptions (such as the distribution of ejected gas or the stellar profiles) differ from the results of the Horizon-AGN run. We therefore find alternative best fit parameters that match our simulation results. The preferred parameters values are $M_c=10^{13.8}$ M$_\odot/h$, $k_s=67$ $h/$Mpc and $\eta_b=0.17$ at this redshift. The lower value of $\eta_b$ compared to the fiducial $\eta_b=0.5$ value of \citet{Schneider15} indicates that the gas ejection of the Horizon-AGN run is less strong than the one assumed in the BC model. This is clearly shown in the comparison to the dashed and dot-dashed curves, which show the impact of changing $\eta_b$ to $0.3$ and $0.5$, respectively, while keeping $M_c$ fixed to the best fit value. Our results also prefer a lower value of $M_c$ compared to the fiducial $10^{14.07}\,$M$_\odot/h$ value adopted in \citet{Schneider15}. This finding is related to the fractional content of gas in haloes, which is discussed in Section \ref{sec:discuss}. The increase in $k_s$, on the other hand, compared to the fiducial $55$ $h/$Mpc value corresponds to a higher concentration of matter at small scales.

The right panel of Figure \ref{fig:bcmcomp1} shows that the extrapolation of the $z=0$ best fit case (in orange) predicts too little suppression of power at $k$ of a few $h/$Mpc. This suggests that the BC model as formulated in Section \ref{sec:bcm} cannot describe the redshift evolution of the Horizon-AGN predictions. To account for a redshift evolution different than originally parametrised, we fit the BC model parameters to the simulation results at each redshift. The preferred parameter values are shown in Figure \ref{fig:bcmcomp2} as a function of redshift. We restrict to the redshift range $0\leq z\leq 3.8$. This high redshift restriction is imposed due the impact of resolution on the stellar population assembly of galaxies, which requires at least 1 Gyr to converge \citep{Blaizot04}. We have verified that the majority of galaxies affected by AGN feedback at $z=3.8$, namely galaxies above a stellar mass threshold of $10^{9.5}\,{\rm M}_\odot$ \citep{Beckmann17}, have stellar populations older than this threshold. 

We find that the parameter $M_c$, which determines the amount of suppression in the total matter power spectrum and is related to the mass of the haloes responsible for it, is roughly constant in the range $0<z<2$, and it increases to higher redshift. The value of $k_s$, which represents the scale below which the contribution of the stellar profile sets in, decreases from $z=4$ to $z=2$, and then starts to increase again towards $z=0$. Finally, $\eta_b$, which sets the physical scale for the suppression of power, also shows a non-monotonic evolution, with a peak at $z=2$. The latter is a consequence of the redshift behaviour of the Horizon-AGN run discussed in Section \ref{sec:cv} which is not captured by the BC model.

The values of these preferred parameters can be interpreted in the context of the BC model as follows. The typical mass of haloes where a large portion of the gas gets ejected by AGN processes is roughly constant between $0\leq z<2$. In Section \ref{sec:results1}, we drew similar conclusions based on our results from previous work \citep{Beckmann17}. We suggested that, despite the availability of more massive haloes at low redshift, AGN feedback is not strong enough once in ``maintenance'' mode. At the same time, $\eta_b$ decreases, restricting the impact of AGN on the total matter power spectrum to smaller and smaller scales and once again indicating a progressively waning impact of AGN on the total matter power spectrum. It is interesting to note that there is an anti-correlation between $k_s$ and $\eta_b$ as a result of the fits, shown in Figure \ref{fig:anticorr}. This is not built into the BC model, but can potentially suggest avenues for reducing its parameter space, as we discuss below.

Figure \ref{fig:bcmcomp2} also shows that the preferred value for $M_c$ increases by several orders or magnitude between $z=2$ and $z=4$. This high redshift increase of $M_c$ effectively reduces the number of haloes which can drive a suppression of power in the BC model. This is accompanied by a decrease of $\eta_b$ in this redshift range, suggesting that gas ejection is less efficient than in the fiducial model considered in \citet{Schneider15}, and the gas is expelled up to shorter distances from the centre of haloes. A non-monotonic evolution of $k_s$, increasing from $z=2$ to $z=4$ is related to the effect of adiabatic cooling, which is already present as early as $z=4.9$ in Figures \ref{fig:PScomp_noAGN_DM} and \ref{fig:PScomp_AGN_DM}.

The fiducial implementation of the BC model does not reproduce the redshift evolution of the Horizon total matter power spectra. We have found that an alternative with a small number of parameters which captures the Horizon results in the range $0\leq z \leq 3.8$ is given by the following parametrisation of $F(k,z)$:
\begin{equation}
  F(k,z)=\frac{1+[k/\kappa_s(z)]^2}{1+[k/\kappa_s(z)]^\beta},
  \label{eq:newpar}
\end{equation}
where $\kappa_s(z)=\kappa_{s,0}+\kappa_{s,1}z+\kappa_{s,2}z^2$ and $\kappa_{s,i}$ for $i=\{0,1,2\}$ are free parameters. The numerator of equation (\ref{eq:newpar}) captures the small scale enhancement in power due to adiabatic cooling, while the denominator models the suppression due to AGN feedback. The typical scale for both effects is connected to the value of $\kappa_s(z)$ as a consequence of the anti-correlation found between BC model parameters in Figure \ref{fig:anticorr}. The preferred values for the parameters are: $\beta=1.39$, $\kappa_{s,0}=28.5$, $\kappa_{s,1}=-11.9$ and $\kappa_{s,2}=2.50$. For these parameters, $F(k,z)$ reproduces Horizon results within $<5\%$ for all redshifts and scales considered. Physically driven modifications to the BC model to account for redshift evolution are not straightforward. This would require looking into the individual components of the model directly in the hydrodynamical simulation, which is the topic of our future work.

\begin{table}
  \begin{center}
    \title{}
    \begin{tabular}{c | c c c c c} 
      \hline
      $i$ & $A_i$ & $B_i$ & $C_i$ & $D_i$ & $E_i$  \\ [0.5ex] 
      \hline
      2  & $80.6$ & $0.86$ & $-0.11$ & $0.10$ & $-2.05$ \\ 
      1  & $-35.1$ & $-1.29$ & $0.67$ & $-0.16$ & $3.60$ \\
      0  & $5.24$ & $0.71$ & $1.45$ & $0.08$ & $1.11$ \\ [1ex] 
      \hline
    \end{tabular}
    \caption{Best fit parameters for $F(k,z)$ in the parametrisation proposed by \citet{Harnois15}, as described by equation (\ref{fkzjhd}).}
    \label{tab:jhd15fit}
  \end{center}
\end{table}

Other effective parameterisations of the impact of baryons on the matter power spectrum have been proposed in the literature. For example, in the work by \citet{Harnois15}, an effective fitting function was proposed to model the impact of baryons on the total matter power spectrum from the OWLS simulations. The functional form proposed by \citet{Harnois15} was ($b_m$ in their notation):
\begin{equation}
  F(k,z)= 1-A_ze^{(B_z x - C_z)^3}+D_z x e^{E_zx},
  \label{fkzjhd}
\end{equation}
where $x=\log_{10}(k/[h\,{\rm Mpc}^{-1}])$, the function $A_z$ is parametrised as 
\begin{equation}
  A_z = A_2a^2+A_1a+A_0,
\end{equation}
with $a$ the scale factor, and similarly for $B_z$, $C_z$, $D_z$ and $E_z$. We find that this expression for $F(k,z)$ is sufficiently flexible to fit the Horizon results to within $3\%$ across $z\leq 3.8$ with preferred parameters as listed in Table \ref{tab:jhd15fit}. The disadvantages of this approach are the large number of free parameters ($15$ in comparison to $3$ in the fiducial BC model and $4$ in equation \ref{eq:newpar}) and the fact that there is no physical interpretation for them. Nevertheless, we make the parameters for this fitting function available here, as it might be useful for comparison to \citet{Harnois15}.

\subsubsection*{Publicly available power spectra}

The power spectra obtained from the Horizon set and used in this manuscript are made publicly available. These correspond to the redshifts labelled in Figure \ref{fig:PScomp_AGN_noAGN} up to $z=3.8$, following the convergence criterion discussed above. Shot noise is subtracted. 

To test the accuracy of the four-parameter fitting function $F(k,z)$ of equation (\ref{eq:newpar}), we obtained additional curves to those presented in Figure \ref{fig:PScomp_AGN_DM} at $z=0.76$ and $z=1.5$. We find that $F(k,z)$ reproduces those results within $3.4\%$ and $4.3\%$, respectively. On the other hand, directly interpolating the set of public power spectra with a cubic spline results in uncertainties of up to $1.5\%$ and $2.2\%$ for the fractional impact of baryons on the total matter power spectrum at $z=0.76$ and $z=1.5$, respectively. The individual power spectra (with and without baryons) are interpolated with much larger uncertainty ($\sim 8\%$) than their ratio.

\section{Discussion}
\label{sec:discuss}

The results presented in this work suggest that there is significant dispersion between predictions of the impact of baryons on the total matter power spectrum from different cosmological hydrodynamical simulations (Figure \ref{fig:compare}) at $z=0$. EAGLE shows the smallest impact from baryons on the distribution of matter at $k<10\,h/$Mpc, while the original Illustris run is most impacted. In addition, simulations also predict different redshift evolution of this effect. Distinguishing between these different scenarios is impossible from simulation data. 

Observational constraints can be used to distinguish between different predictions. Numerous observations are already available that test the accuracy of the hydrodynamical simulations across the redshift range of future surveys. Those observations include galaxy luminosity and mass functions \citep[see a comparison of Horizon-AGN results to these observations in][]{Kaviraj17} or stellar to halo mass relations \citep{Leauthaud12,Wojtak13,Velander14,Han15,vanUitert16}, for example. 

\begin{figure}
\includegraphics[width=0.47\textwidth]{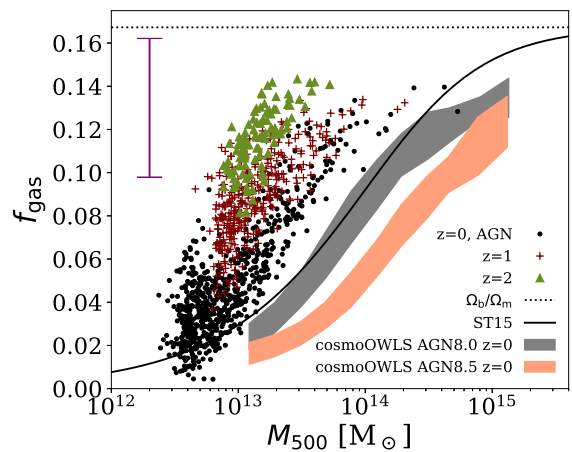}
\caption{The fraction of gas by mass in haloes, $f_{\rm gas}$, as a function of the mass enclosed within $r_{500}$. The dotted line represents the baryon fraction for our adopted cosmology. The solid line is a parametrisation of $z=0$ observational constraints presented in \citet{Schneider15}. The black, red and green dots correspond to Horizon-AGN results at redshifts $z=0$, $z=1$ and $z=2$, respectively. The shaded area indicate results from the cosmo-OWLS simulations \citep{LeBrun14}, specifically: the gray area corresponds to the AGN8.0 model (similar to the one adopted in the OWLS AGN run) and the orange area, to the AGN8.5 model, with increased heating temperature in the AGN feedback implementation. The purple error bar in the top left indicates the typical scatter of current $f_{\rm gas}$ constraints from \citet{Gonzalez13}.}
\label{fig:fgas}
\end{figure}

In their work, \citet{Schneider15} used constraints on the fraction of bound gas within haloes to inform the BC model. Similarly, one can compare how well the simulations reproduce this specific observation, as we show in Figure \ref{fig:fgas}. Haloes in Horizon are identified using {\sc Adaptahop} \citep{aubertetal04}. This halo finder identifies a local density as a halo if it has more than $50$ particles and if density computed from the twenty nearest neighbours exceeds 178 times the cosmological average density. For each halo in Horizon-AGN we have extracted the fraction of the mass in gas within the volume enclosing a sphere of $500$ times the critical density of the Universe, and we show it as a function of $M_{500}$, the total mass of the halo within the corresponding radius ($r_{500}$). Figure \ref{fig:fgas} shows that the Horizon-AGN simulation over-predicts the fraction of gas in massive haloes (black points) with respect to the parametrised observational results at $z=0$ described by the black solid line \citep{Schneider15}. This is the cause for our lower preferred value of $M_c$ in the BC model fits of Section \ref{sec:effmod}, shown in Figure \ref{fig:bcmcomp1}. The excess gas fraction in haloes suggests a lower strength of AGN feedback, which in the BC model is associated with a lower halo mass.

The shaded areas in Figure \ref{fig:fgas} show the fraction of gas at $z=0$ in the haloes of the cosmo-OWLS simulations \citep{LeBrun14}. The grey area corresponds to the AGN8.0 model, which shares the same parameters as the OWLS `AGN' run. The orange area corresponds to the AGN8.5 cosmo-OWLS model, which adopts a slightly increased heating temperature for AGN feedback \citep{Booth09}. Using synthetic X-ray observations to estimate the halo mass and gas mass fraction of simulated clusters, \citet{LeBrun14} found that, in a {\it WMAP}7 cosmology, observational results lie in between the predictions of the cosmo-OWLS AGN8.0 and AGN8.5 runs. The red crosses and green triangles in Figure \ref{fig:fgas} represent Horizon-AGN results at $z=1$ and $z=2$, respectively. The fraction of gas inside haloes decreases towards $z=0$. Nevertheless, the decrease is not sufficient to bring the Horizon-AGN predictions into agreement with observations, and the simulation tends to over-predict the amount of gas in haloes. While cosmo-OWLS provides a better fit to the gas fraction, notice that the scatter in current observational constraints is large, as represented by the purple error bar shown in that figure, taken from \citet{Gonzalez13} as an example. Cosmo-OWLS also shows some discrepancies with observations in respect to the stellar content of halos \citep{LeBrun14}.

We emphasise that while some mitigation strategies for the impact of baryons on the matter power spectrum and cosmological observables from galaxy surveys have been proposed \citep{Semboloni11,Semboloni13,Harnois15,Mead15}, these have often relied on the OWLS predictions from \citet{vanDaalen11} alone, the first predictions on the impact of baryons on the total matter power spectrum to become publicly available. \citet{Eifler15} is the exception, using the OWLS results in combination with two other sets of simulations. It is known that the success of these mitigation techniques depends on the ability to cover the parameter space of possible models \citep{Mohammed17}. Our results suggest that hydrodynamical simulations can give very different predictions for the amplitude-, redshift- and scale-dependence of the effect. In principle, the success of the available mitigation strategies should be tested against the different cosmological simulations taking into account observational priors at the same time. An alternative is to perform a more aggressive calibration of sub-grid parameters based on observational data. \citet{McCarthy17} have recently reported on a new suite of simulations ({\sc BAHAMAS}) which adopt similar sub-grid models as cosmo-OWLS and where the free parameters are calibrated to reproduce the present-day stellar mass function and gas fraction in groups and clusters. In \citet{McCarthy18}, it was shown that current uncertainties in these low redshift observations are sufficiently small to provide useful information for cosmological purposes.

\citet{Semboloni11} pointed out potential degeneracies between the impact of baryons on the total matter power spectrum and other physical processes. These included estimates of neutrino mass, the running of the spectral index to constrain inflation and the nature of dark matter. In this regard, \citet{Villaescusa17} recently suggested that the impact of neutrinos can be isolated from that of baryonic processes because of the distinct scale- and redshift-dependence of the effects. On the other hand, other degeneracies remain unexplored, including degeneracies with other astrophysical and observational systematics that affect weak lensing (photometric redshifts, source blending, intrinsic galaxy alignments). In particular, \citet{Hearin12} showed that increased uncertainties in photometric redshifts can make requirements on knowledge of the power spectrum more stringent.

Finally, considering that the statistical uncertainties on the total matter power spectrum from future weak lensing surveys (percent level) will be much smaller than the effect of baryons on this observable, it is expected that weak lensing measurements in the next decade will be able to distinguish between different AGN feedback models \citep{Foreman16}, although this depends on the level of control over weak lensing systematics at small scales. 

\section{Conclusions}
\label{sec:conclude}

In this work, we presented results on the distribution of matter at cosmological scales from the Horizon set of simulations in terms of the total matter power spectrum between $z=0$ and $z=5$. We found that intermediate scales ($1<k<10\,h/$Mpc) are suppressed by $5-15\%$ with respect to the DMO run. At smaller scales, the power spectrum is enhanced due to efficient cooling and star formation. The Horizon-noAGN simulation also displays suppression at high redshift compared to the DMO run, associated to the additional pressure provided by the baryons. At low redshifts, Horizon-noAGN is purely enhanced with respect to the DMO run. We have also quantified the impact of baryons on the distribution of dark matter in the simulation, finding it to be of a few percent at low redshift. The results are subject to cosmic variance and the presence of massive haloes, which we have verified by splitting the simulation volume into $8$ sub-volumes. 

A comparison between the Horizon set and other cosmological hydrodynamics simulations (Section \ref{sec:comparesims}) shows that the impact of baryons is smaller in Horizon, of typically $\sim 12\%$ at a scale of $k\sim10\,h/$Mpc at $z=0$. Correspondingly, the fraction of gas inside haloes in the Horizon-AGN simulation is enhanced compared to low redshift observations, suggesting AGN feedback as implemented here is not strong enough to eject or prevent the infall of material into haloes (Figure \ref{fig:fgas}). In this regard, the cosmo-OWLS AGN8.0 run is in better agreement with observations although the scatter in current observational constraints of $f_{\rm gas}$ is large. It is also possible that our over-predicting gas fractions is a late consequence of the weak stellar feedback. As large halos assemble hierarchically from smaller ones, if the smaller halos contain an excessive amount of gas at the time of the merger, an overly strong AGN would be required to compensate. This nuanced approach to making the halos have gas fractions in better agreement with observations is consistent with \citet{Spacek2017}, who find that merely increasing the AGN feedback efficiency would make the thermal Sunyaev-Zeldovich signal for large-mass halos even less consistent with observations. The investigation of how sub-grid models interplay in producing the simulated gas fractions is unfortunately not possible without running additional simulations. In this context, we note that additional exploration of the impact of the minimum heating temperature on the distribution of gas is needed. \citet{Hahn17} have suggested that AMR simulations are not very sensitive to the choice of this parameters, in comparison to smoothed-particle-hydrodynamics simulations \citep{LeBrun14}. Nevertheless, their results from a suite of hydrodynamic AMR zoom simulations of massive clusters are not fully representative of the specific heating temperature and cosmological volume of Horizon-AGN.

The redshift evolution is also different between simulations, with Horizon displaying a non-monotonic trend associated with the impact of gas pressure and AGN feedback, which competes with the growth of structure between $0<z<6$. It is likely that choosing different sub-grid parameters for the feedback efficiency could lead to modified predictions in this redshift evolution. The Horizon results are well-approximated by the BC model of \citet{Schneider15} at $z=0$, but with different preferred values for the parameters that determine the amount and scale of suppression of power due to baryons, and a smaller typical scale for the stellar component. The redshift dependence of our results also differs from the BC model prediction and we have provided an effective parametrisation with $4$ free parameters which approximates the Horizon results within $<5\%$ at all redshifts. In the future, we plan to carry out a more detailed comparison between the BC model assumptions and components and the simulation predictions.

The total matter power spectra obtained in this work from Horizon-AGN, Horizon-noAGN and Horizon-DM have been made publicly available.

\section*{Acknowledgments}
This work has made use  of the HPC resources of CINES (Jade and Occigen supercomputer) under the time allocations 2013047012, 2014047012 and 2015047012 made by GENCI. This work is partially supported by the Spin(e) grants {ANR-13-BS05-0005} (\url{http://cosmicorigin.org}) of the French {\sl Agence Nationale de la Recherche} and by the ILP LABEX (under reference ANR-10-LABX-63 and ANR-11-IDEX-0004-02). We thank S. Rouberol for running  smoothly the {\tt Horizon} cluster for us. Part of the analysis of the simulation was performed on the DiRAC facility jointly funded by STFC, BIS and the University of Oxford.

NEC acknowledges support from a Beecroft Postdoctoral Research Fellowship and a Royal Astronomical Society Research Fellowship. RSB acknowledges support from STFC. AMCLB was supported by the European Research Council under the European Union's Seventh Framework Programme (FP7/2007-2013) / ERC grant agreement number 340519.

We thank the anonymous referee for comments that helped improve this manuscript. We are grateful to Volker Springel, Ruediger Pakmor, Mark Vogelsberger, Lars Hernquist and Wotjek Hellwing for providing the matter power spectra from IllustrisTNG, Illustris and EAGLE for comparison to Horizon-AGN in this work. We thank the OWLS team for making their results publicly available. We are grateful to Joachim Harnois-D\'eraps for useful discussions and to Robert J. Thacker for help setting up and running {\tt pispec4} for the power spectrum computation. Some of the theoretical predictions used in this work have made use of the Core Cosmology Library\footnote{\url{https://github.com/LSSTDESC/CCL}}, which also uses the {\tt CLASS} software \citep{CLASS}.

\bibliographystyle{mn2e_warx}
\bibliography{author}


\appendix
\section{Accuracy at large scales}
\label{app_growth}

Several of the results presented in this manuscript  (Figs. \ref{fig:PScomp_AGN_noAGN} through \ref{fig:bcmcomp1}) refer to ratios between power spectra of Horizon-AGN, Horizon-noAGN and Horizon-DM, the three simulations of the Horizon set. In taking these ratios, it is often the case that the redshifts being compared across runs differ slightly. For example, one of the snapshots is extracted at redshift $z_1$, while the other one is extracted at $z_1+\delta z$\footnote{The time-step in the Horizon runs is set by the Courant condition: $\delta t =\mathcal{C} \delta x/v_{\rm max}$, where $\Delta x$ is the spatial resolution of the AMR grid, $\mathcal{C}=0.8$ is a constant factor, and $v_{\rm max}$ is the maximum characteristic velocity in the simulation (considering both fluid and particles). $v_{\rm max}$ is usually set by the maximum sound speed, and the stronger the feedback (i.e. the presence or not of AGN), the higher its value. As a consequence, this results in different time-steps for the different runs.}. To first order, we can and we do correct for this effect by re-scaling the power spectra using the linear growth function. In other words, in presenting the ratio of power spectra from simulations A (at $z_1$) and B (at $z_1+\delta z$), we estimate
\begin{equation}
\frac{\Delta^2_A(z_1)}{\Delta^2_B(z_1)}=\frac{D^2(z_1+\delta z)}{D^2(z_1)}\frac{\Delta^2_A(z_1)}{\Delta^2_B(z_1+\delta z)},
\end{equation}
where $D$ is the linear growth factor normalised to $1$ at $z=0$. Even after performing this correction, we have found residual effects that affect the ratio of power spectra across simulations at large scales at the $1\%$ level, which can be clearly seen in Figures \ref{fig:PScomp_AGN_noAGN} and \ref{fig:PScomp_AGN_DM} at $z=0$ for example. The re-scaling procedure based on the linear growth function is insufficient to remove the large-scale excess power.

On the one hand, we know that even the largest scales in the box are growing nonlinearly by $z=0$. We have verified that this is the case by performing the following comparison. For each of the simulation runs, we compared power spectra at two different redshifts by taking their ratio at large-scales normalised by the growth factor such that
\begin{equation}
  \chi \equiv \frac{D^2(z_2)}{D^2(z_1)}\frac{\Delta^2_A(z_1)}{\Delta^2_A(z_2)}.
  \label{eq:lscomp}
\end{equation}
In practise, we chose $z_1$ and $z_2$ to be consecutive snapshots among those presented in Figure \ref{fig:PScomp_AGN_noAGN}. The comparison demonstrated that indeed equation (\ref{eq:lscomp}) deviates from unity by $\sim 1\%$ at large scales. Based on the results of \citet{Schneider16}, who studied the convergence of the large-scale power in $N$-body simulations with different box sizes, this result is expected. \citet{Schneider16} indeed suggest that a minimum volume of $(500\,h/{\rm Mpc})^3$ is required for the simulation to probe linear scales at $z=0$. 

\begin{figure}
\includegraphics[width=0.47\textwidth]{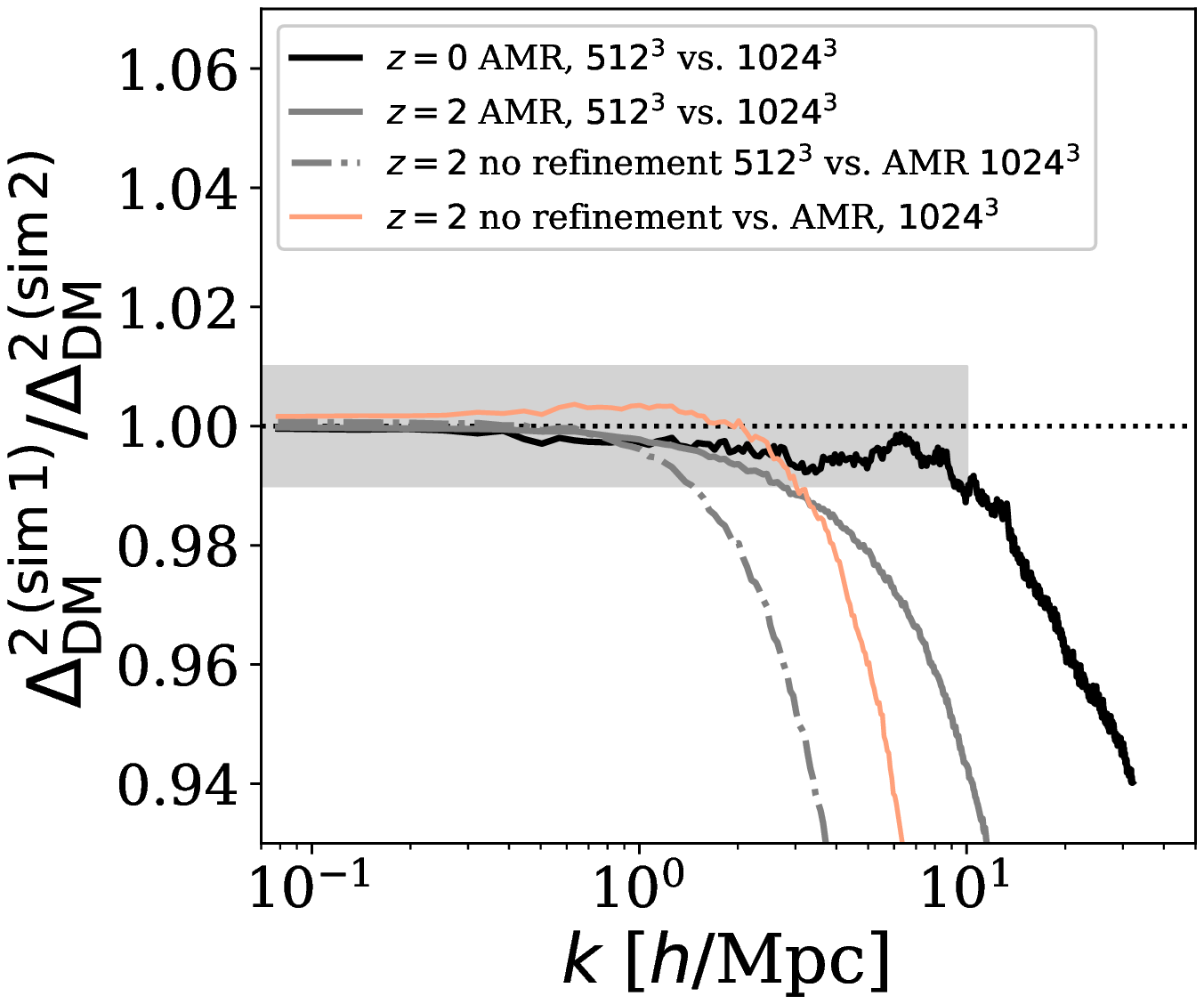}
\caption{The impact of resolution (i.e., number of particles) in the matter power spectrum of the DMO run. The figure shows the ratio between the power spectrum from different low resolution runs and our fiducial run with $1024^3$ particles. The black curve shows the results for the run with $512^3$ particles and refinement at $z=0$; the grey curve corresponds to the same simulation at $z=2$. The dot-dashed curve corresponds to the $z=2$ case without AMR. The orange curve corresponds to an unrefined run with $1024^3$ particles at $z=2$. The shaded area represents $\pm 1\%$ accuracy.}
\label{fig:res}
\end{figure}

We also considered the possibility of a transfer of power from small to large-scales due to the refinement scheme of the RAMSES code. We ruled out this hypothesis by comparing the matter power spectrum from the Horizon-DM simulation at $z=2$ to that of an additional unrefined run with $1024^3$ particles extracted at the same redshift. We show this comparison in the orange curve of Figure \ref{fig:res}, which demonstrates that any large-scale difference in power is much less than $1\%$ in this case.

There are alternative possible explanations for the $1\%$ excess of power at large scales. The work by \citet{Angulo13} has demonstrated that depending on the gravitational kernel, the coupling of baryons and DM can result in spurious large-scale power even in the case of linear theory. Further, this phenomenon is sensitive to the way that the initial conditions are set for the baryons and the dark matter \citep{Valkenburg17}. Other hydrodynamic simulations seem to be affected by small offsets in large-scale power as well (see Figure 13 of \citealt{Mummery17}). We emphasise that a $1\%$ accuracy at large scales satisfies our requirements for this work.

\section{Convergence tests}
\label{app_res}

In this section, we study the convergence properties of the matter power spectrum at small scales in the Horizon simulation. The fiducial resolution of Horizon is $1024^3$ particles, with an approximate dark matter mass of $8\times 10^7$ M$_\odot$ in the baryonic runs, and slightly higher in the DMO run to accommodate the same $\Omega_{\rm m}$ value. To estimate the convergence rate of the simulation, we run two other DMO boxes with $512^3$ particles (with and without refinement) and a $1024^3$ box without refinement, and we compare the matter power spectrum estimated from those boxes to the fiducial one in this work at $z=2$ and $z=0$. 

The results are shown in Figure \ref{fig:res}. The shaded area in the figure represents the target of $1\%$ accuracy below $k<10\,h/$Mpc. The results for $z=2$ (solid grey) indicate that, due to resolution effects on the number of particles, this accuracy is achieved at $k<3\,h/$Mpc. The additional impact of the refinement of the grid is then evidenced in the comparison between the solid grey curve and the dot-dashed grey curve, which corresponds to the unrefined $512^3$ simulation at $z=2$. The reduction in the convergence scale defined by the $1\%$ accuracy requirement is of a factor of $\sim 2$ for the dot-dashed grey curve. Focusing now on the orange curve which indicates the results of the $1024^3$ unrefined run at $z=2$, we find that $1\%$ accuracy is achieved at $k=3\,h/$Mpc. Extrapolating from the $512^3$ case, this suggests that the $1024^3$ refined simulation should have achieved $1\%$ convergence at $k\sim 6\,h/$Mpc at least. Its convergence rate compared to a hypothetically $2048^3$ refined simulation should be shallower than in the case of the orange curve, thus suggesting we very likely achieve a few percent convergence throughout all the scales of interest in this work.

At $z=0$, the convergence rate is much better than at $z=2$. This is evidenced from the results of the black solid curve, which shows the comparison of the $512^3$ refined simulation to the $1024^3$ refined simulation at $z=0$. In this case, the convergence scale is extended to $k\sim 10\,h/$Mpc, thus allowing us to infer that the $1024^3$ runs have converged to approximately twice that value. Note that the convergence rate in the case of the presence of baryons should be even better due to the increased number of particles (roughly a factor of $2$ at $z=0$).

Our results are in good agreement with \citet{Schneider16}, who studied the accuracy of matter power spectrum predictions from DMO simulations with a variety of box sizes and resolutions. Their results suggested that per cent accuracy can be achieved up to $k\sim 4\,h/$Mpc for a $(512$ Mpc$/h)^3$ simulation with $1024^3$ particles. Re-scaling the wave-number to our box size, we expect to achieve per cent accuracy at around $k\sim 20\,h/$Mpc, in line with the result quoted in the paragraph above. We conclude that we have achieved the desired numerical convergence in the matter power spectrum, and as the main results of this work highlight, the impact of baryons on the total matter power spectrum exceeds the numerical accuracy at the scales of interest and depends on the exact implementation of baryonic physics (Figure \ref{fig:compare}).

\end{document}